\documentclass[5p,review,number,sort&compress]{elsarticle}

\usepackage{lineno}
\modulolinenumbers[5]

\usepackage[T1]{fontenc}
\usepackage{datetime}
\usepackage{graphicx}
\usepackage{color}

\usepackage{bm}
\usepackage{multirow} 
\usepackage{amsmath}
\usepackage{dcolumn}
\usepackage{textcomp}
\usepackage{epstopdf}
\usepackage{booktabs}
\usepackage{calc}
\usepackage[percent]{overpic}
\usepackage[version=3]{mhchem} 
\newcommand*{\vcenteredhbox}[1]{\begingroup
\setbox0=\hbox{#1}\parbox{\wd0}{\box0}\endgroup}

\journal{Nuclear Instruments and Methods in Physics Research A}

\begin{document}

\begin{frontmatter}

\title{Feasibility and Applications of the Spin-Echo Modulation Option for a Small Angle Neutron Scattering Instrument  at the European Spallation Source \tnoteref{t1}}
\tnotetext[t1]{\copyright\enskip 2017. This manuscript version is made available under the CC-BY-NC-ND 4.0 license http://creativecommons.org/licenses/by-nc-nd/4.0/  }

\author{A. Kusmin}
\author{W. G. Bouwman}
\author{A. A. van Well}
\author{C. Pappas}
\address{Department of Radiation Science and Technology, Faculty of Applied Sciences, Delft University of Technology, 2629JB Delft, The Netherlands}




\begin{abstract}
We describe theoretical and practical aspects of spin-echo modulated small-angle neutron scattering (SEMSANS)  as well as the potential combination
with SANS.  Based on the preliminary technical designs  of SKADI (a SANS instrument proposed for the  European Spallation Source) and a SEMSANS add-on, 
we assess the  practicability, feasibility  and scientific merit of a combined SANS and SEMSANS setup by calculating  tentative SANS and SEMSANS results for soft matter, geology and advanced material samples that have been previously studied by scattering methods. We conclude that lengths from  1 nm up to  0.01 mm can be observed simultaneously in a single measurement.
Thus, the combination of SANS and SEMSANS instrument  is suited for the simultaneous observation of a wide range of length scales, e.g. for time-resolved studies of kinetic processes in complex multiscale systems.
\end{abstract}


\end{frontmatter}

%

\section{\label{sec:intro}Introduction}
The range of length scales typically observed in a small-angle neutron scattering (SANS) experiment is between 1 nm and several 100 nm. 
 Larger length scales on the $\mu$m range can be observed either with ultra SANS (USANS)  \cite{BarkerAgamalian2005USANSNIST}  or with spin-echo small-angle neutron scattering  (SESANS)  \cite{RekveldtPlompBlaauw2005SESANSDELFT}.
Recently, a new technique to measure SESANS  using the modulation of neutron spin-echo polarization across the incident beam has been suggested and tested
 \cite{BouwmanDuifGaehler2009Larmor,StrobBouwmanlPlompTOFSEMSANS2012,StroblWiederBouwman2012SEMSANS,SalesStrobl2015SEMSANS,StroblSalesHabicht2015SEMSANSNature}.
As the components required to perform the modulation are located before the sample and do not affect the configuration of the SANS instrument, 
 the idea of a combined spin-echo modulation small angle neutron scattering (SEMSANS) and a SANS instrument has been put forward  \cite{BouwmanDuifGaehler2011SESANS}. 
 
 An implementation of this idea was proposed for SKADI, one of 
 the SANS instruments to be built at ESS, and has been included in the list of potential add-ons for this instrument  \cite{JakschBouwmanFrielinghaus2014SKADI}.
Preliminary requirements and potential designs of the SEMSANS add-on
 have been discussed in the technical reports  \cite{DuartePintoESSDelivWP111,DuartePintoESSDelivWP112}.
 In this paper, we take a detailed look into the technical feasibility of combined SANS and SEMSANS measurements and the potential applications.
 

We start by a short description of the SEMSANS technique, and the relation between the SANS and SEMSANS results. After that, based on the
technical designs of SKADI and of a SEMSANS add-on, we calculate the boundaries of length scale regions accessible by SANS  and SEMSANS and show that they overlap.
We discuss the impact of the SEMSANS add-on on SANS measurements, and calculate SEMSANS signals   
 for soft matter, geological,  advanced material samples that have already been studied by neutron scattering. The results of the calculations 
 show that the combination of SANS with SEMSANS is feasible and  can cover a wide range of length scales  simultaneously,  which makes it useful for 
 kinetic process studies in complex, multiscale samples in time resolved mode.

\begin{figure*}[htp]
\centering
\includegraphics[width=0.9\textwidth]{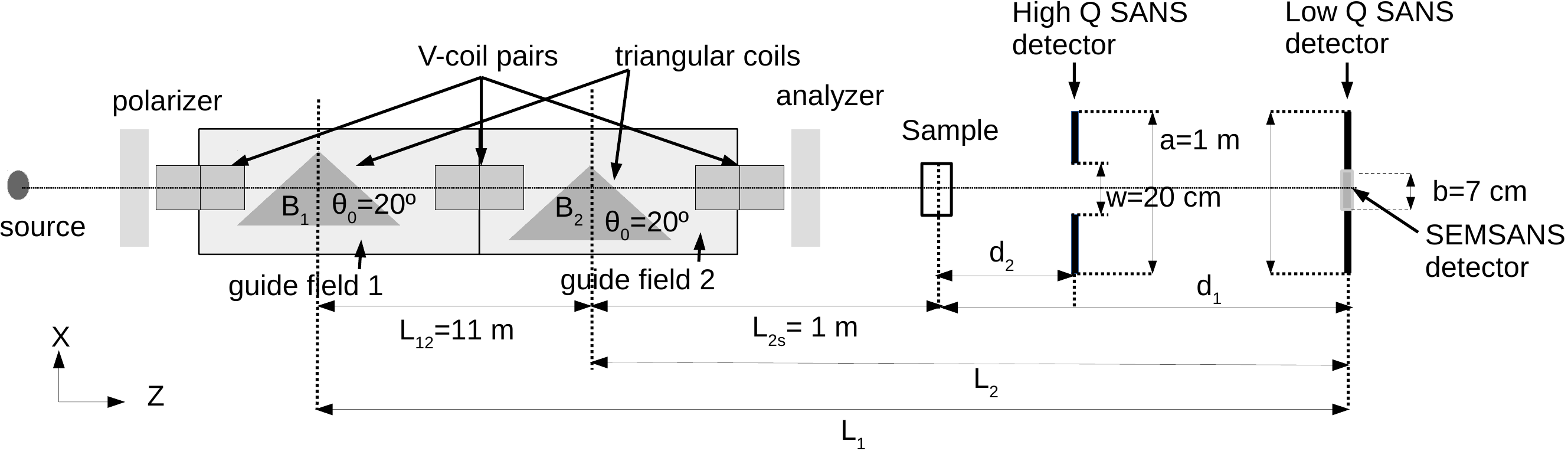}
    \caption{A sketch of a SANS+SEMSANS instrument (view from above, not to scale). The magnetic fields $B_1$ and $B_2$, 
    and guide fields 1 and 2 are parallel to $Y$ axis and point into one direction. The first and the last V-coil pairs act as $\pi/2$-flippers, the middle pair acts as a $\pi$-flipper.
Both SANS detectors are squares\,  $a \times a$; the high $Q$ SANS detector 
has a square window in the center, $w \times w$. 
The beam stop is a square, $b \times b$.
    The size of the SEMSANS detector is limited by the beam stop size.
The distance $d_2$ is fixed at $0.2 \times d_1$, $d_1$ can be set to 8 m or 20 m. 
    The angle $\theta_0=20^{\circ}$ corresponds to triangular  coils tested in Delft  \cite{SalesStrobl2015SEMSANS}.
}\label{fig:semsansscheme}
\end{figure*}%

\section{\label{sec:technique} A SEMSANS experiment}
\subsection{\label{subsec:thetechnique} The technique}
A simplified scheme of the SANS instrument together with a SEMSANS add-on is  shown in Fig.  \ref{fig:semsansscheme}. 
The neutron beam is polarised in the $Y$ direction. 
The first pair of V-coils  \cite{KraanRekveldtPor1991} acts as a $\pi$/2-flipper: it adiabatically 
rotates the polarisation into the $XZ$ plane  thus triggering Larmor precessions. The central pair of V-coils performs a  $\pi$-flip, and the last pair stops the precessions by performing a $\pi$/2-flip.
The net precession angle is the difference between the precession angles $\phi_1$ and $\phi_2$ accumulated before and after the  $\pi$-flip, respectively.
 The spin echo (se) polarization measured after transmission through an analyser  is defined as: $P_{se}=P_0\cos(\phi_1-\phi_2)$, 
where  $P_0$ is the beam polarization in the absence of Larmor precessions, typically the product of the  polarizer and analyzer efficiencies, and $\phi_i=\gamma B_i t_i$,
where $\gamma$ is neutron gyromagnetic ratio,
 and $t_i$  is the time-of-flight of the neutron beam through the $i$th triangular region\footnote{For brevity, the contribution of the $i$th guide field  to $\phi_i$ is omitted. This is justified because the guide fields are tuned in such a way that  their contribution to the net precession angle is zero.}.


Let us first assume that all neutron trajectories are parallel to $Z$ axis. 
The  middle of the first triangular region is at $x=0$, the path through this region is $ l_1(x)=\frac{H -2x}{\tan \theta_0}$, where $H$ is the height of the triangle.
The middle of the second  triangular region is shifted towards negative $x$ by $\Delta$, the path through this region is $ l_2(x)=\frac{H -2(x+\Delta)}{\tan \theta_0}$.

 Using $t_i(x)=l_i(x) m \lambda/h$, where $h$ is Planck constant, $m$ and $\lambda$ are neutron's mass and wavelength, 
and Larmor constant, $c=m\gamma/ h$, we arrive at:
\begin{subequations}
    \begin{align}
     \phi_1 =  [ c \lambda B_1 H/ \tan \theta_0 ] - [2 c \lambda B_1 x  / \tan \theta_0 ] 
    \label{eq:precessionangle1a}\\
     \phi_2 =  [ c \lambda B_2 (H -2 \Delta) / \tan \theta_0 ]  - [2 c \lambda B_2 x / \tan \theta_0] 
    \label{eq:precessionangle1b}
    \end{align}
\end{subequations}
The shift $\Delta$ is fixed at 
\begin{equation}
\Delta=H(1-B_1/B_2)/2
    \label{eq:shiftdelta}
\end{equation}
to make  $\phi_1-\phi_2$=0 at $x=0$, that is, to realize spin-echo in the center of the detector.

From  eqs. \ref{eq:precessionangle1a}-\ref{eq:precessionangle1b} the spin echo polarization is 
\begin{equation}
    P_{se}(x)=  P_0\cos(2\pi x/\zeta )
    \label{eq:precessionangle2}
\end{equation}
where the oscillation period, $\zeta$, is given by:
\begin{equation}
    \zeta =  \frac{ \pi \tan \theta_0 }{c \lambda (B_2-B_1)} 
    \label{eq:zeta}
\end{equation}

The resulting neutron intensities for the eigenstates $|+\rangle$ and $|-\rangle$ 
are:
\begin{equation}
    I_{se}^{\pm}(x)=\frac{I(x) [1 \pm P_{se}(x) ]}{2} 
    \label{eq:SEMSANSintensity}
\end{equation}

So far we neglected the dependence of the net precession angle on the  divergence in the $XZ$ plane.
For a divergent beam, 
 when the distances between the detector, and the centers of the first and the second triangular regions, $L_1$ and $L_2$, respectively,  
 satisfy the condition \cite{BouwmanDuifGaehler2009Larmor}:
\begin{equation}
    B_1 L_1 =B_2 L_2 
    \label{eq:B1L1B2L2}
\end{equation}
 measured intensities can be described by a modified version of eq. \ref{eq:SEMSANSintensity}:
\begin{equation}
    I_{se}^{\pm}(x)=\frac{I(x) [1 \pm V(x)\cos(2\pi x/\zeta ) ]}{2}
    \label{eq:SEMSANSintensityNEW}
\end{equation}
where  $V(x)$ is the so-called visibility.

In the absence of a sample, the visibility  is $V_0(x)$:
\begin{equation}
    V_0(x)= P_0 R_{pixel}R_{\Delta \lambda}(x)
    \label{eq:SEMSANSvis}
\end{equation}
where  
$ R_{\Delta \lambda}(x)$ accounts for the effect of wavelength resolution (see Appendix A), and 
$ R_{pixel}$ accounts for a finite spatial resolution of the detector (see Appendix B).

In the presence of a sample, 
the $x$-coordinate of the pixel where a neutron will be detected depends on whether it is scattered or not and, if it is scattered, on the scattering angle.
Thus, the visibility of modulations will be reduced to
an extent that depends on the intensity and  angular dependence of the small-angle scattering, more details are given in the next section.
The normalized visibility, $V_s=V(x)/V_0(x)$, thus reflects the scattering from the sample only, because the
 normalization cancels all additional effects due to the finite wavelength resolution and spatial detector resolution.

Please note that the shift of the second triangular precession region relative to the first one, i.e. $\Delta$  from  eq. \ref{eq:shiftdelta}, is 
independent of magnetic field  settings because of the condition expressed by eq.  \ref{eq:B1L1B2L2}. 
The use of such a shift is a new approach.
 Alternatively, instead of triangular coils, magnets with an inclined foil flipper can also be used \cite{BouwmanDuifGaehler2011SESANS}.

%
%
%
\subsection{\label{subsec:equations} Relation between SEMSANS, SESANS,  and SANS }
It was  shown in Ref. \cite{Strobl2014SEMSANSSCIREP} that  
the normalized visibility, $V_s$, is the equivalent of the normalized neutron spin-echo polarization 
measured by SESANS. Therefore, it can be described by the same  
theory  \cite{RekveldtKreuger2003SESANSandCoMSC} leading to:
\begin{equation}
    V_s(\delta_{SE}) = \exp[\Sigma_{exp} t (G_{exp}(\delta_{SE})-1)]]
    \label{eq:pz}
\end{equation}
where $t$ is the sample thickness, and the subscript $exp$  reflects the dependence of respective parameters on experimental conditions (such as detector size, sample-detector distance etc).   
In SEMSANS the spin-echo length is given by the relation:
\begin{equation}
    \delta_{SE}=\lambda d_1/\zeta  =\frac{c   (B_2 - B_1)d_1 \lambda^{2} }  {\pi \tan \theta_0 } 
    \label{eq:spinecholengthSEMSANS}
\end{equation}
where $d_1$  is the distance between sample and detector. 

For small scattering angles\footnote{That is, when $\tan \theta \approx \theta$, and $Q_z \approx 0$. This condition is fulfilled for SEMSANS.} $\Sigma_{exp}$ is related to  the differential coherent SANS cross-section $\frac{d\Sigma(Q)}{d\Omega}$, through:
\begin{equation}
    \Sigma_{exp} = \frac{\lambda^2}{(2 \pi)^2 }\int_{pixel}  \frac{d\Sigma(\bm{Q})}{d\Omega}\mathrm{d} Q_x \mathrm{d}  Q_y 
    \label{eq:scatcrossection1}
\end{equation}
The meaning of $\Sigma_{exp}$ can be best seen for the case of a thin sample, in which case the product $\Sigma_{exp}t$  is just the fraction of neutrons that are scattered into the solid angle covered by the pixel of the SEMSANS detector. 

The function $G_{exp}(\delta_{SE})$ in eq. \ref{eq:pz} is the experimentally observed  projection of the autocorrelation function of the scattering length density,
$\gamma(\bm{r})$ \cite{AnderssonBouwman2008JAC}.  Its theoretical definition, $G_{tot}(\delta_{SE})$, can be expressed 
 as the Hankel transform of $d\Sigma(Q)/d \Omega$ if the scattering is isotropic  \cite{AnderssonBouwman2008JAC}:
\begin{equation}
    G_{tot}(\delta_{SE}) = \frac{1}{2\pi\xi_{tot}}\int_0^{\infty} J_0(Q\delta_{SE}) \frac{d\Sigma(Q)}{d \Omega}  Q \mathrm{d} Q 
    \label{eq:GZfromIQhankel}
\end{equation} 
where $J_0$ is the zeroth-order Bessel function of the first kind, and the subscript $tot$ (for "total") 
indicates that the upper integration limit is infinity.
The correlation length,  $\xi_{tot}$, is given by  \cite{AnderssonBouwman2008JAC}:
\begin{equation}
    \xi_{tot} =  \frac{1}{2\pi}\int_0^{\infty}  \frac{d\Sigma(Q)}{d \Omega}  Q \mathrm{d} Q 
    \label{eq:defineksi}
\end{equation} 

The experimentally determined $V_s(\delta_{SE})$  can be analyzed using eq.  \ref{eq:pz}
  where the functions 
 $\Sigma_{exp}$ and $G_{exp}(\delta_{SE})$ can be related to $d\Sigma(Q)/d \Omega$ through: 
\begin{equation}
    \xi_{exp} = \frac{\Sigma_{exp}}{ \lambda^2} = \frac{1}{2\pi}\int_0^{Q^{max}_{SEMSANS}}  \frac{d\Sigma(Q)}{d \Omega}  Q \mathrm{d} Q  
    \label{eq:defineksiexp}
\end{equation} 
\begin{equation}
    G_{exp}(\delta_{SE}) = \frac{1}{2\pi\xi_{exp}}\int_0^{Q^{max}_{SEMSANS}} J_0(Q\delta_{SE}) \frac{d\Sigma(Q)}{d \Omega}  Q \mathrm{d} Q 
    \label{eq:GZfromIQhankelexp}
\end{equation} 
The obvious difference between  eqs. \ref{eq:GZfromIQhankel}, \ref{eq:defineksi}  and eqs. \ref{eq:defineksiexp}, \ref{eq:GZfromIQhankelexp} is in the integration range, which in the latter case is limited by acceptance angles to the maximum accessible $Q$, $Q^{max}_{SEMSANS}$. 

A detailed account for an effect of acceptance angles on measured SEMSANS intensities and on $V_s(\delta_{SE})$ has been given in Appendix C. 
In case when  $Q^{max}_{SEMSANS}$ is particularly low, that is, when a significant part of small-angle scattering cross-section is not registered by the SEMSANS detector,  eq.  \ref{eq:pz} no longer holds and  a more general result, eq.  \ref{eq:newapp2} in Appendix C, should be used. 

\begin{table*}[thb]
    \caption{SKADI instrument configurations and the limits on the accessible $Q$   and spin-echo length ($\delta_{SE}$)  ranges. 
        For calculation details see 
     Sec. \ref{sec:skadithing}.
     ${\delta_{SE~B}^{min}}$ and  ${\delta_{SE~B}^{max}}$ are the limits  due to minimum and maximum magnetic fields,
     respectively. 
    ${\delta_{SE~pixel}^{max}}$ is the limit due to the spatial resolution of the SEMSANS detector. 
 }  
 \small
\label{table1SKADI}
\center\begin{tabular}{l|c|c|c|c|c|c|c}
\toprule
\hline
\multicolumn{3}{c|}{Configuration} &\multicolumn{2}{c|}{SANS range} &\multicolumn{3}{c}{SEMSANS range  }  \\
\hline
    Mode   & $d_1$ [m] &  $\lambda^{min}$;$\lambda^{max}$ [\AA]     &  $Q_{SANS}^{min}$; $Q_{SANS}^{max}$ [\AA\textsuperscript{-1}]  & 
$\frac{2\pi}{Q_{SANS}^{min}}$   [$\mu$m]  &
    ${\delta_{SE~B}^{min}}$ [$\mu$m] & ${\delta_{SE~B}^{max}}$ [$\mu$m]  & ${\delta_{SE~pixel}^{max}}$ [$\mu$m]      \\
\midrule
\hline
    High flux & 20  & 3; 8.5 & 1.8 $\times$ 10\textsuperscript{-3}; 0.37 & 0.34      & 0.038 &30 & 126 \\
    Wide $Q$ & 20  & 3; 14 & 1.1 $\times$ 10\textsuperscript{-3};  0.37 & 0.57       & 0.038 & 82 & 207 \\
    High flux& 8 & 3; 10.2 & 3.8 $\times$ 10\textsuperscript{-3};  0.87 & 0.16    & 0.036 & 28 & 60 \\
    Wide $Q$ & 8 &3; 17.4 & 2.2 $\times$ 10\textsuperscript{-3};  0.87 & 0.28       &0.036 & 81 & 103 \\
\hline
\bottomrule
\end{tabular}
\end{table*}%
%
%
%
%
%
\section{\label{sec:skadithing} Practical aspects of combining SANS and SEMSANS}

In order to assess the impact of  the SEMSANS add-on on a SANS instrument, it is important to estimate the accessible $Q$- and $\delta_{SE}$-ranges and identify a possible overlap. This was performed by fixing the major instrument parameters, for SANS according to the technical design of SKADI  \cite{JakschBouwmanFrielinghaus2014SKADI}, and for SEMSANS according to previous reports  \cite{DuartePintoESSDelivWP111,DuartePintoESSDelivWP112}.  

\subsection{\label{subsec:sansthing}   $Q$-range accessible to SANS}
In the high flux mode SKADI uses every neutron pulse leading to a bandwidth of 5.5 \AA{} and 7.2 \AA{} for a sample-detector distance of $d_1$=20 m and $d_1$=8 m, respectively.  
In the wide $Q$ mode every second pulse is used leading to a doubling of the bandwidth.  
For the combination with SEMSANS, the incident neutron beam must be polarized; therefore, based on the current SKADI design, the minimum wavelength $\lambda^{min}$ cannot be lower than 3 \AA{}  \cite{JakschBouwmanFrielinghaus2014SKADI}. 


The minimum $Q$ for SANS is given by:
\begin{equation}
    Q^{min}_{SANS}=\frac{4\pi}{\lambda^{max}}\sin(\mathrm{atan}(\frac{0.5\sqrt{2}b}{d_1})/2)\approx \frac{\sqrt{2}\pi b}{d_1 \lambda^{max}}
    \label{eq:qminsans}
\end{equation}
where $b$ is the size of the beam stop. Since  $d_2 = 0.2 \, d_1$,  the maximum $Q$ accessible by SANS is: 
\begin{equation}
    Q^{max}_{SANS}=\frac{4\pi}{\lambda^{min}}\sin(\mathrm{atan}(\frac{0.5\sqrt{2}a }{0.2d_1})/2) \approx \frac{5\sqrt{2}\pi a}{d_1 \lambda^{min}}
    \label{eq:qmaxsans}
\end{equation}
where $a$ is the size of the SANS detector.
\subsection{\label{subsec:semsansthing}   $\delta_{SE}$-range accessible to SEMSANS}
The spin-echo length range can be calculated from  eq. \ref{eq:spinecholengthSEMSANS} and  eq. \ref{eq:B1L1B2L2}.
Since the measurements can be done at more than one sample-to-detector distances ($d_1$)  and, therefore, 
at different $L_1$ and $L_2$ distances, it 
 is preferable  to  make the $d_1$-dependence explicit and rewrite  these equations as follows:
\begin{equation}
    \delta_{SE}=\frac{d_1 L_{12} c   B_1 \lambda^{2} }  {(d_1+L_{2s})\pi \tan \theta_0 } 
    \label{eq:spinecholengthSEMSANSrepeat2}
\end{equation}
\begin{equation}
    B_2=B_1\frac{L_{12}+L_2}{L_2} =B_1\frac{L_{12}+L_{2s}+d_1}{L_{2s}+d_1} 
    \label{eq:b1max}
\end{equation}
where $L_{2s}$ and $L_{12}$ are as defined in Fig. 1. 

The following parameters are fixed: $\theta_0=$ 20\textdegree,  $L_{2s}$=1 m, and $L_{12}=11$ m. 
The other two parameters, $d_1$ and $\lambda$, depend on the configuration of the host SANS instrument.  Once the latter is fixed, 
the $\delta_{SE}$ range can only be modified  by changing  $B_1$.
The magnetic field limits of the triangular coils are $B_1^{min}$  =0.1 mT and $B_2^{max}$ = 15 mT. 
The limiting $\delta_{SE}$ values are $\delta^{min}_{SE~B}$ (calculated from $\lambda^{min}$ and $B_1^{min}$), and   
 $\delta^{max}_{SE~B}$  (calculated from $\lambda^{max}$ and  $B_1^{max}$, which, in turn, is calculated from $B_2^{max}$ and eq. \ref{eq:b1max}). 

Due to the finite detector spatial resolution the visibility of the sample and empty beam measurements is reduced by a factor $R_{pixel}$, which depends on the pixel size and oscillation period $\zeta$ (see Appendix B). The minimum acceptable $R_{pixel}$ is set to 0.75 and the corresponding limit on the maximum spin-echo length, $\delta^{max}_{SE~pixel}$, is calculated from  eq. \ref{eq:deltaSEmax}  using $\lambda^{max}$ and a pixel size of 55 $\mu$m, a spatial resolution which is already reached by state-of-the art detectors  \cite{TremsinFilges2012SEMSANSDETECTOR}.

As can be seen from  Tab.  \ref{table1SKADI}, the minimum and maximum spin-echo lengths are related to the  
 magnetic field limits,  leading to  the range of $[\delta_{SE~B}^{min} ; \delta_{SE~B}^{max}]$. This range must be distinguished from  the range  covered in a single measurement with an arbitrary $B_1$, which is  
 $[\delta_{SE}(B_1,\lambda_{min});$ $ \delta_{SE}(B_1,\lambda_{max})]$.

The comparison of the maximum lengths observable with SANS 
($2\pi/Q^{min}_{SANS}$) and the minimum lengths observable with SEMSANS ($\delta_{SE~B}^{min}$)  reveals a substantial overlap between the length scales  
accessible to SANS and SEMSANS, for all instrument configurations.
\subsection{\label{subsec:dmin}  The  effect of acceptance angles}
%
%
As it can be seen from  eq. \ref{eq:GZfromIQhankelexp},  the correlation function $G_{exp}(\delta_{SE})$ is 
a  Fourier transform  over an experimentally accessible $Q$-region limited by 
\begin{equation}
    Q^{max}_{SEMSANS} \approx \frac{2\pi\theta^{max}_{SEMSANS}}{\lambda}
    \label{eq:qmaxSEsetup}
\end{equation}
where  $\theta^{max}_{SEMSANS}$ is the maximum accepted scattering angle. 
The  $\theta_{SEMSANS}$-limits  are given by (see Appendix D):
\begin{equation}
    \theta_{SEMSANS}\in ( \frac{x}{d_1} \pm \big( \frac{S}{d_{1}}+ \frac{S_C+S}{d_C} \big) )
    \label{eq:thetamaxSEsetup}
\end{equation}
where $x$ is the $X$-coordinate of the SEMSANS detector pixel, the sizes of sample and collimation apertures along $X$ axis are $2S$, and $2S_C$, respectively, and the distance between the two apertures is $d_{C}$.

For $d_{1}$= 8 m,  $d_{C}$=8 m, $S_C$=1.5 cm, and $S$=0.5 cm (aperture sizes correspond to Fig. 7 in Ref.  \cite{JakschBouwmanFrielinghaus2014SKADI})  
 and for $x=0$,   $|\theta_{SEMSANS}|  < 3.125$ mrad.
With increasing $x$, the range of accepted angles becomes increasingly asymmetric around zero.
For example, at $x=1$ mm, $-3 < \theta_{SEMSANS} < 3.25 $ mrad, 
 at $x$=1.5 cm, $-1.25 < \theta_{SEMSANS} < 5 $ mrad.
This asymmetry decreases at larger  $d_1$-distances.

Because of an $x$-dependence of the accepted $\theta_{SEMSANS}$-range, the function $G_{exp}(\delta_{SE})$ and normalized visibility $V_s({\delta_{SE}})$  may be  $x$-dependent as well (cf. eq. \ref{eq:GZfromIQhankelexp} and eq. \ref{eq:pz}).
In general, a fit of a SANS model to $V_s({\delta_{SE}})$  can account for this $x$-dependence. 
However, for simplicity, such an $x$-dependence can also be removed.
To do that, only the intensity measured in a narrow central region of the SEMSANS detector around $x=0$,  e.g.  $|x|<2$ mm should be used to calculate $V_s(\delta_{SE})$. 
In this paper, we  choose the latter option and calculate  $Q^{max}_{SEMSANS}$, $G_{exp}(\delta_{SE})$, and $V_s(\delta_{SE})$  for $x$=0.

Finally, please note that to increase counting statistics,  $V_s(\delta_{SE})$ can be calculated from measured intensities that are summed up along the $Y$-axis of the SEMSANS detector. In such a case, however, the range of accepted scattering angles in $YZ$ plane will be broader than in $XZ$ plane. This can be taken into account 
by using a modified version of eqs.  \ref{eq:defineksiexp},\ref{eq:GZfromIQhankelexp}, where a single integral over $|Q|$ is replaced by a double integral over $Q_x$ and $Q_y$. More details are given in Appendices C and D. Calculations in this paper were done under assumption that  $Q^{max}=Q_x^{max}=Q_y^{max}$. 

\subsection{\label{subsec:SANSeffect}  Interference of the SEMSANS add-on with the SANS measurements}

As  can be seen in Fig.  \ref{fig:semsansscheme}, all SEMSANS components with the exception of the detector are located before the sample. 
Thus, they do not interfere with the scattered beam. Their impact consists in an attenuation of the incident beam and possibly a modification of the collimation by the polarizer,  V-coils, triangular coils, and the analyzer.

The major impact of a  polarizer and an analyzer on SANS measurements consists in a reduction of the incident beam intensity  by at least factor 2$\times$2=4. 
 If a \textsuperscript{3}He cell is used as  analyzer the divergence of the incident beam is not affected but the intensity losses will be higher.

The scattering from the coils is mainly due to refraction, which is  significant for  cylindrical wires   \cite{PlompBarkervanWell2007}. Therefore, the parts of the coils exposed to the beam should use flat, smooth and thin wires to minimize refraction, absorption and scattering. The neutron beam passes through  six wire planes of the six V-coils at 90\textdegree{} angle and  four wire planes of two triangular coils at 20\textdegree{} angle.  For a thickness of a planar wire of 0.5 mm the total wire thickness becomes $6\times0.5+4\times0.5/\sin(20^{\circ}) \approx$ 9  mm. For aluminium wires,  this corresponds to a transmission of 98\% at $\lambda=$  3 \AA, and 93\% at $\lambda=$ 10 {}\AA.

Behind the sample, a state-of-the-art MCP detector may be used, such as described in ref.  \cite{TremsinFilges2012SEMSANSDETECTOR}. The dimensions of its  active surface are 3 cm by 3 cm and are smaller than dimensions of a beam stop (7  cm by 7 cm), but its actual size is larger. This size will probably be reduced in the future. Alternatively, the MCP detector can be placed behind the high $Q$ SANS detector if the latter has a window in the center.

In addition, as the 
 pixel size of the SANS detector is expected to be  3 mm by 3 mm,
intensity oscillations with a period of a few mm could be observed with SANS as well. 
Their amplitude can be estimated from eq. \ref{eq:SEMSANSvis} by calculating the factor $R_{pixel}R_{\Delta \lambda}(x)$ using eqs.
  \ref{eq:cosy2} and \ref{eq:distancedll2} for the borders of the beam stop, i.e. $x=\pm b/2$.
  For $\zeta=$ 16 mm (the largest period in the results shown later), $R_{\Delta \lambda}(x=b/2,\mathrm{d}\lambda/\lambda=0.04$)$\approx$0.97, and $R_{pixel}(p$=3mm$)$=0.94. Thus, 
  the amplitude of intensity oscillations is almost the same as in the SEMSANS detector. 
  On the other hand, in the case of a smaller  period of 1.6 mm, the oscillations will not be observed by SANS because of $R_{\Delta \lambda}(x=b/2,\mathrm{d}\lambda/\lambda=0.04)\approx$0.06, and $R_{pixel}(p$=3mm$)=$ 0.06. 
  
  The effect of oscillations on the measured SANS cross-sections can be removed by taking the shim or the average of the 
 intensities  (cf. eq. \ref{eq:SEMSANSintensityNEW}).   In any case no oscillations will be observed with the high $Q$ SANS detector because the condition in eq.  \ref{eq:B1L1B2L2} will never be fulfilled.  


%
\subsection{\label{subsec:quality}  Practical aspects with respect to the analysis of SANS and SEMSANS measurements}

A substantial difference in the combined SANS-SESANS data analysis arises from multiple scattering effects, which affect differently the measured patterns. In SANS, where  multiple scattering is a major issue, two extreme cases can be considered.

In the case of multiple small-angle scattering, the probability to be scattered at a small angle is much higher than at larger angles. As a result, the shape of a measured SANS curve is affected, however, procedures such as described in Ref. \cite{ScheltenSchmatz1980}
allow to correct for this and to obtain accurate structural information.

In the second case, multiple scattering  is  primarily caused by the high probability to be scattered at angles larger than the angles accepted by a SANS detector, e.g. due to a substantial incoherent scattering cross-section. In this case, multiple scattering does not change the $Q$-dependence of the measured SANS cross-section, but  it leads to a 
 scattering background, which, however, can be corrected using empirical procedures   \cite{ShibayamaNagao2009SANS}. 

In the following we will focus on multiple SANS effects, as they are the most frequent. Their impact can be evaluated by calculating the SANS transmission, $T_{SANS}(\lambda)$, from eq. \ref{eq:transscat}. This impact is substantial for  $T_{SANS}(\lambda)$  lower than 80\%  (see Appendix E for more details).

In SEMSANS, multiple scattering is taken into account by eq. \ref{eq:pz}, however, it becomes a severe limitation when it leads to $V_s(\delta_{SE})$  close to zero.
In such a case multiple scattering  
limits the maximum usable spin-echo length, which becomes  lower  than $\delta_{SE}^{max}$-limits given in Tab. \ref{table1SKADI}.
To prevent this, we define a constraint, $T_{SEMSANS}>0.01$, where
\begin{equation}
    T_{SEMSANS}=\exp(-\Sigma_{exp}t)
    \label{eq:tsemsans}
\end{equation}
 and $\Sigma_{exp}$  is given by eq. \ref{eq:defineksiexp}.
 For $0<G(\delta_{SE})<1$ 
 it follows from eq. \ref{eq:pz} and eq. \ref{eq:tsemsans}  that  $T_{SEMSANS}< V_s(\delta_{SE})<1$.

For an accurate test of scattering models against experimental SEMSANS data using  eq. \ref{eq:pz}, 
 we estimate  that $V_s(\delta_{SE})$ should be lower than 0.99, which  is somewhat arbitrary but  based on experience with SESANS experiments. Qualitatively, 
 it can also be justified as follows: for  a relative error of 1\%    the deviation of $V_s(\delta_{SE})$ from unity becomes significant only for $V_s(\delta_{SE})$  less or equal to 0.99.

\begin{figure}[htpb]
\centering
    \vcenteredhbox{
    \begin{overpic}[scale=.25]{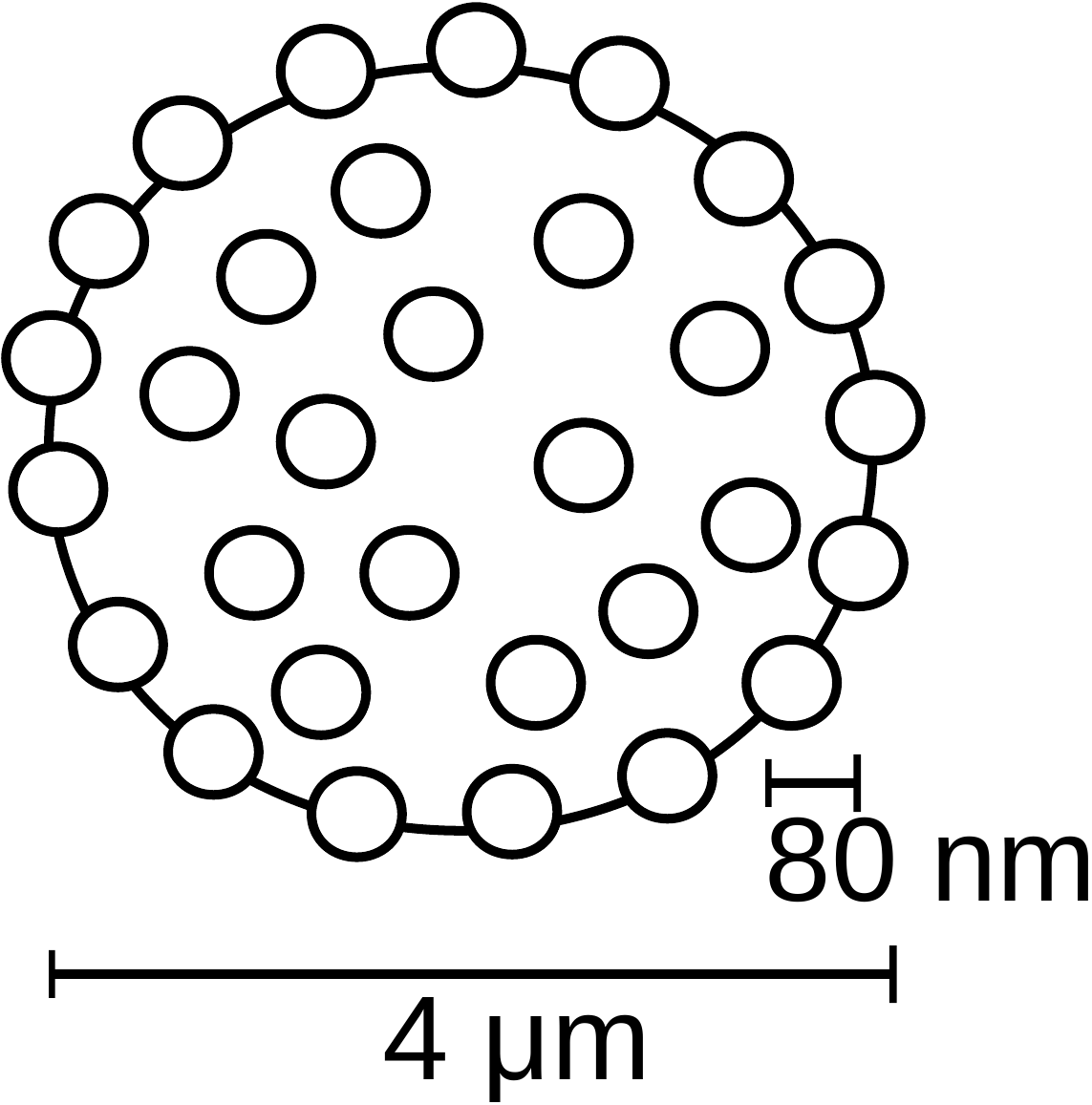}
        \put(50,105){\bf{(A)}}
\end{overpic}
}
\vcenteredhbox{
    \begin{overpic}[scale=.3]{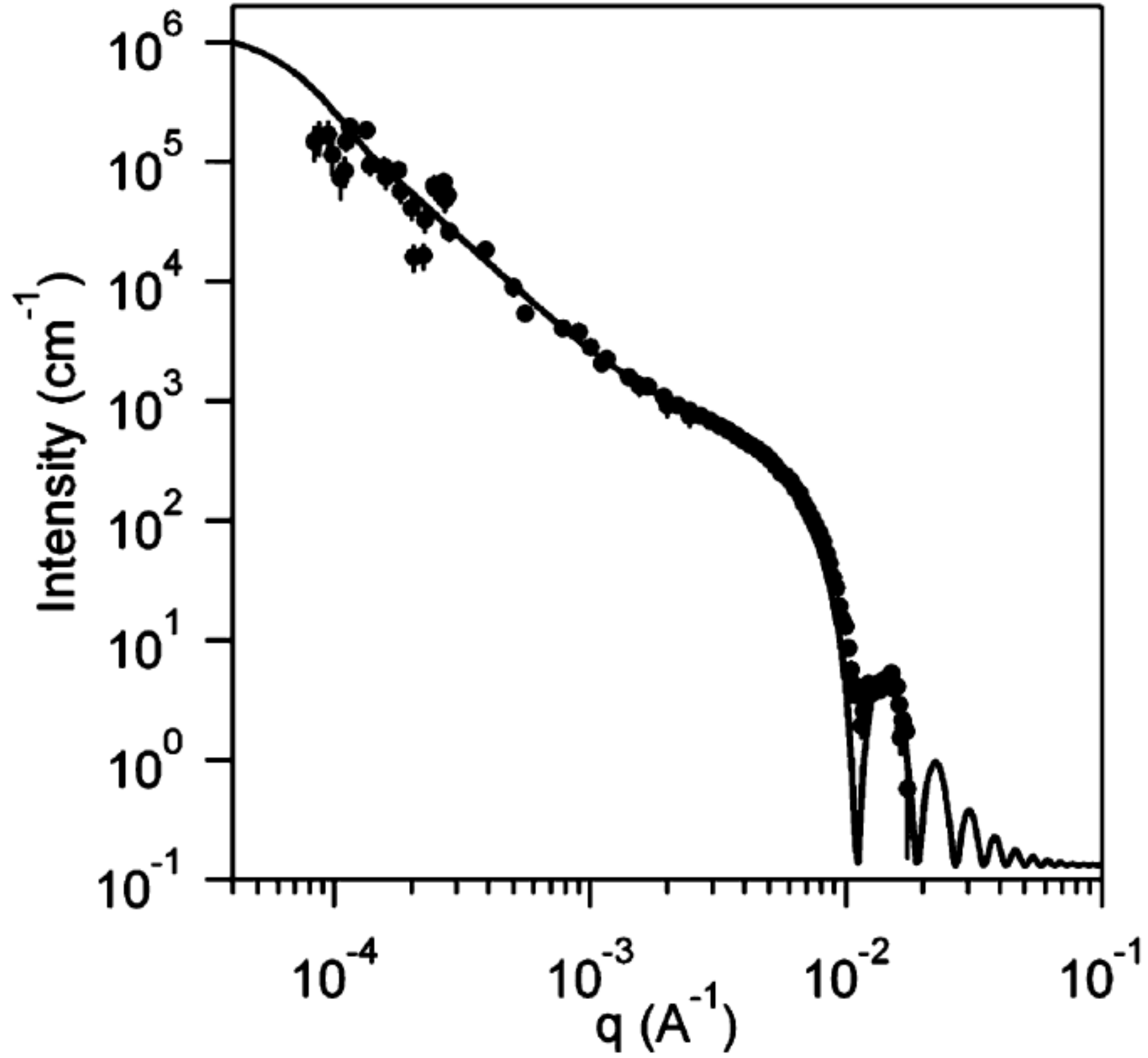}
        \put(75,75){\bf{(B)}}
\end{overpic}
}
\caption{A: Sketch of a raspberry particle. 
    B: Measured and calculated differential SANS+USANS cross-sections; adapted with permission from ref.  \cite{Larson-SmithJacksonPozzo2012Raspberry}; copyright (2012) American Chemical Society.
}\label{fig:raspberry1}
\end{figure}
\begin{figure*}[ht]
\centering
\vcenteredhbox{ \begin{overpic}[scale=.34]{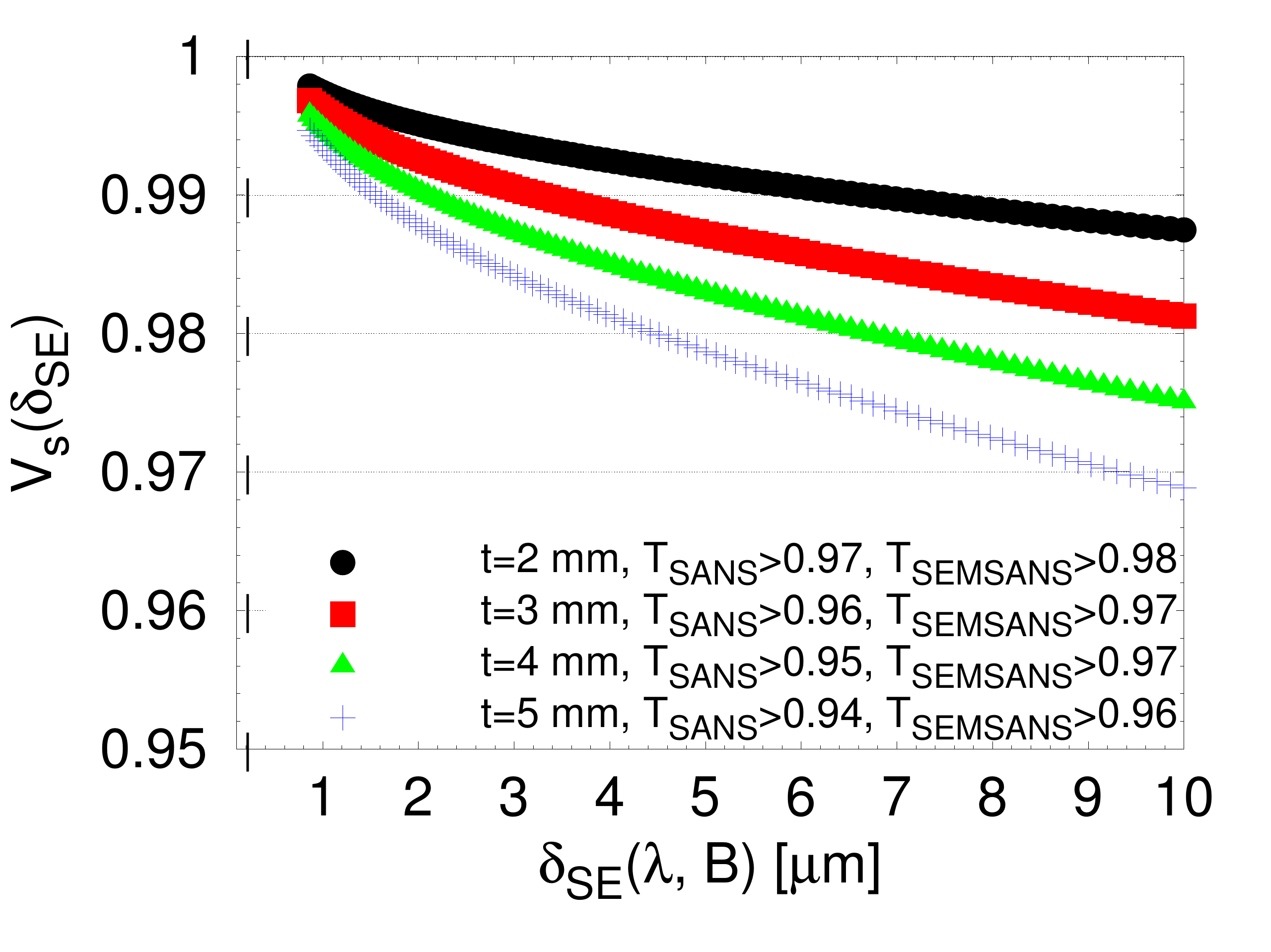}
        \put(75,36){\bf{(A)}}
\end{overpic} }
\vcenteredhbox{ \begin{overpic}[scale=.34]{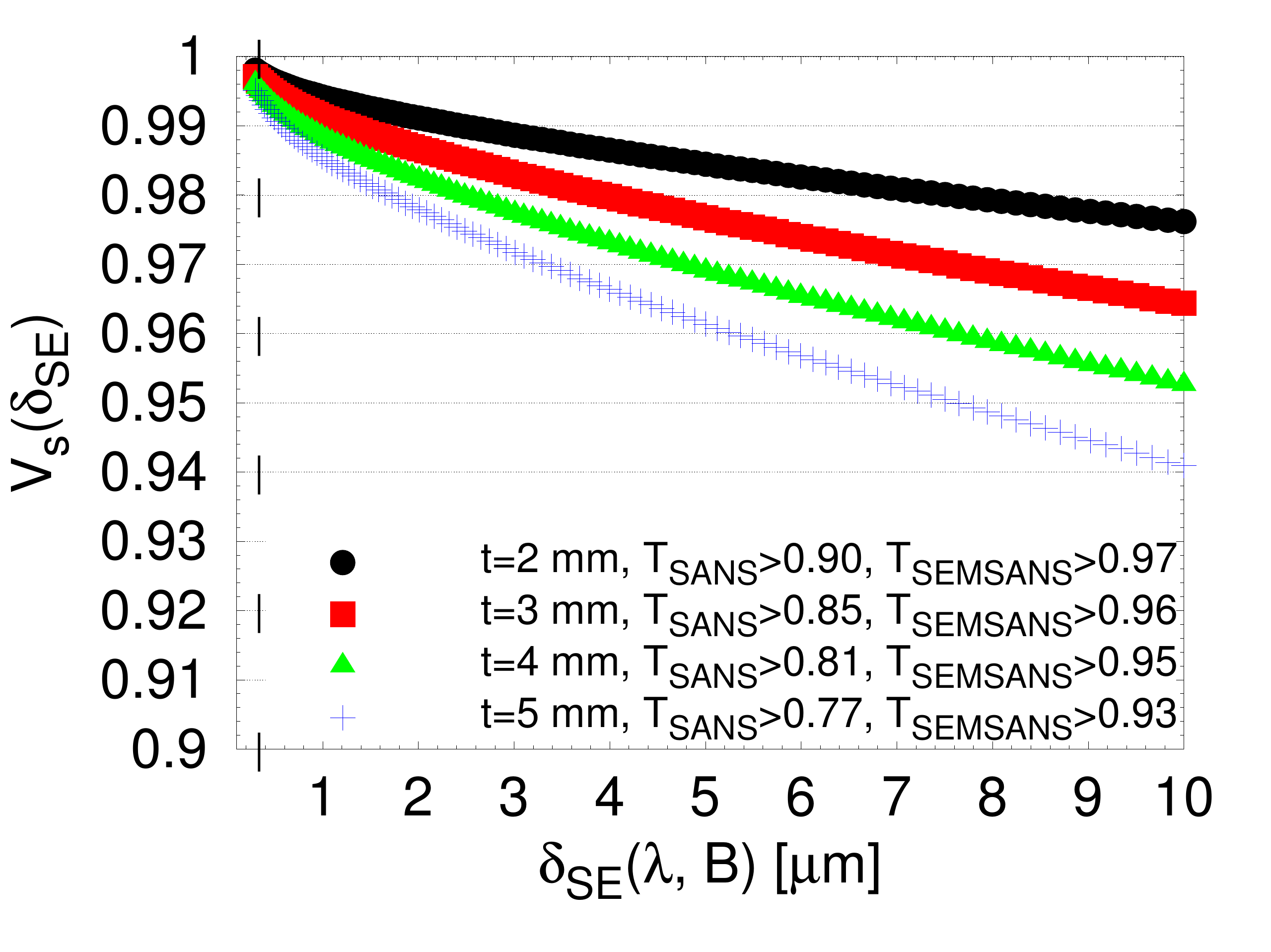}
        \put(75,65){\bf{(B)}}
\end{overpic} }
\vcenteredhbox{ \begin{overpic}[scale=.34]{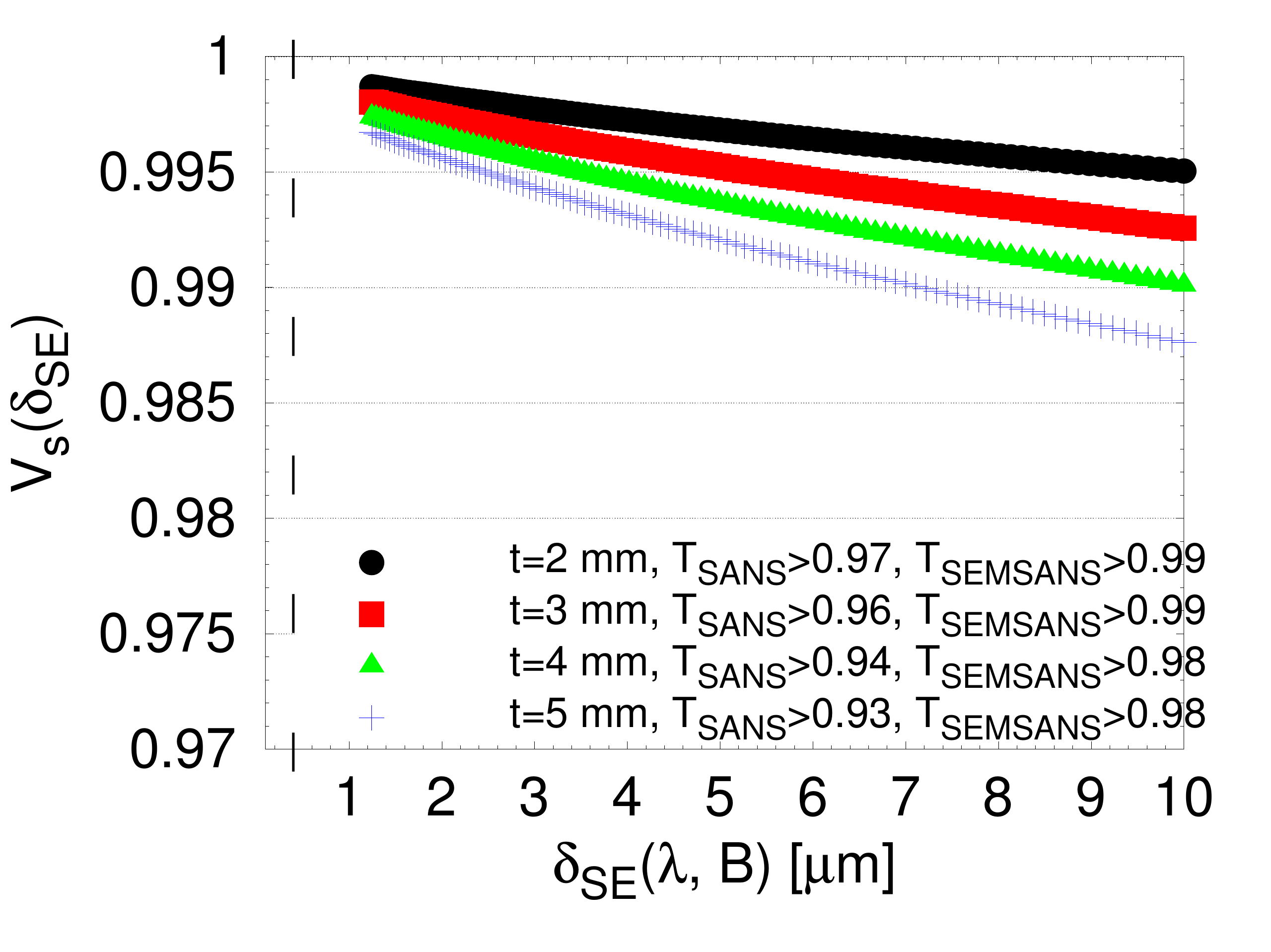}
        \put(75,41){\bf{(C)}}
\end{overpic} }
\vcenteredhbox{ \begin{overpic}[scale=.34]{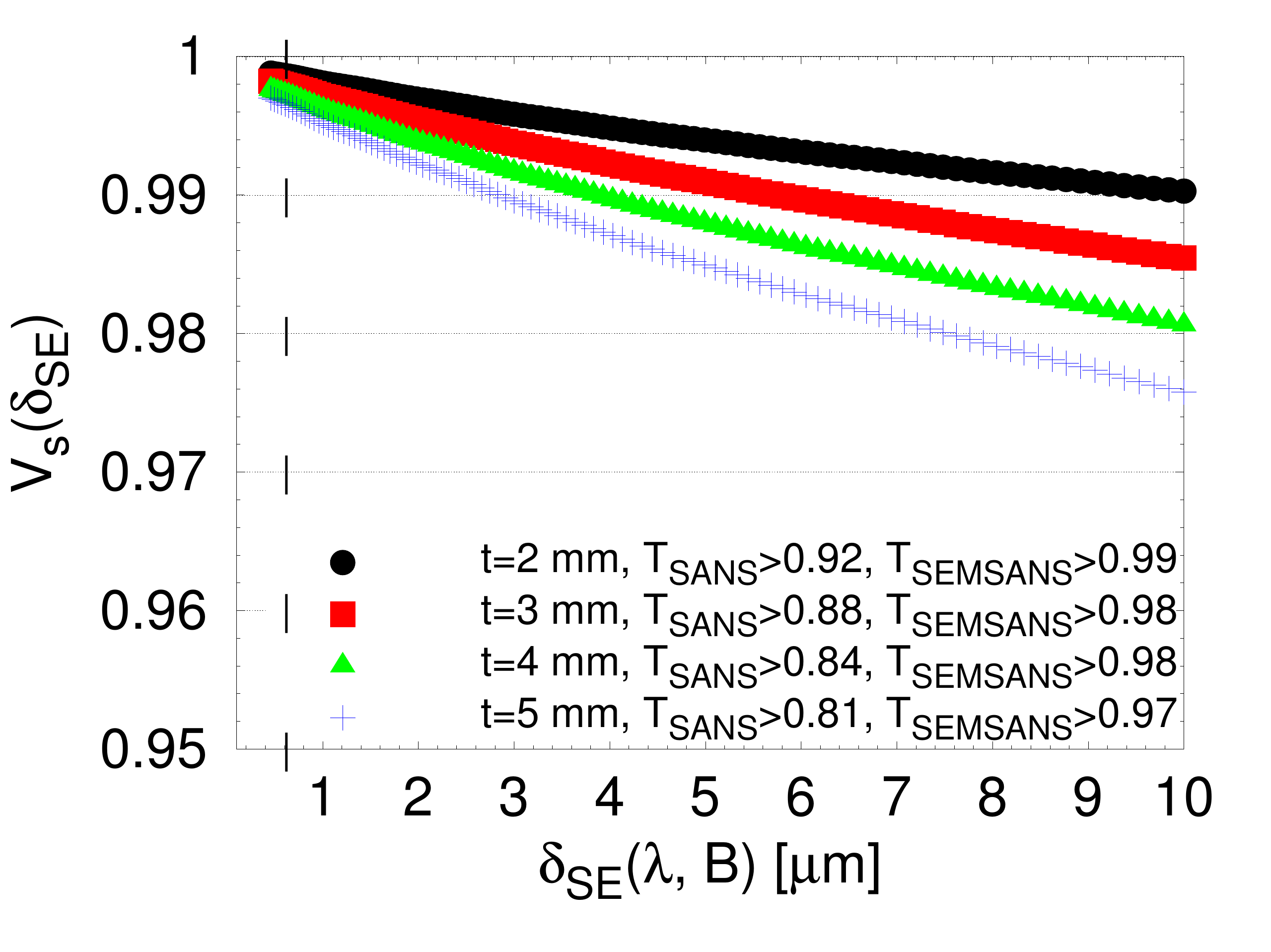}
        \put(75,41){\bf{(D)}}
\end{overpic} }
\caption{$V_s(\delta_{SE})$ for raspberry particles, at several thicknesses ($t$),  $\lambda_{min}=3$ \AA. A: high flux mode, $d_1$=8 m,  $\lambda_{max}=$ 10.2 \AA, $B_1$=2.4 mT, $\zeta^{max}=$2.8 mm; B: wide $Q$ mode, $d_1$=8 m,  $\lambda_{max}=$ 17.4 \AA, $B_1$=0.8 mT, $\zeta^{max}=$8 mm; C: high flux mode, $d_1$=20 m, $\lambda_{max}=$ 8.5 \AA, $B_1$=3.3 mT, $\zeta^{max}=$5 mm; D: wide $Q$ mode, $d_1$=20 m,  $\lambda_{max}=$ 14 \AA, $B_1$=1.2 mT, $\zeta^{max}=$13 mm. 
    A dashed vertical line corresponds to $\delta_{SE}=2\pi / Q^{min}_{SANS}$, that is, the maximum length probed by SANS.
}
\label{fig:raspberrypolz1}
\end{figure*}%

\subsection{\label{subsec:calculationscheme}    Calculation of SEMSANS signal and sample transmission}

As mentioned above several samples have been considered and for every one, we first calculated  $d\Sigma(Q)/d \Omega$ using the model and parameters 
from the original publication (see the  supplementary material for details). Then we selected one of the four SKADI configurations and the corresponding minimum  wavelength, $\lambda^{min}$. As a result, the total wavelength range was fixed to ($\lambda^{min}$,  $\lambda^{max}$), and the $Q$-range accessible with SANS was fixed as well.
The maximum $\delta_{SE}$, $\delta_{SE}^{max}$, was chosen based on the maximum (anticipated) size of structural features in the sample. 
 Using 
eq. \ref{eq:spinecholengthSEMSANSrepeat2} and $\lambda^{max}$, we calculated  $B_1$ and 
 the entire   $\delta_{SE}$-range.
 Then, the following calculations were done in the specified order:
\begin{enumerate}
    \item   $\xi_{tot}(\delta_{SE})$ from eq.  \ref{eq:defineksi};
\item   $G_{tot}(\delta_{SE})$ from  eq. \ref{eq:GZfromIQhankel};
\item the acceptance angle $\theta^{max}_{SEMSANS}$ from eq. \ref{eq:thetamaxSEsetup} calculated for the center of SEMSANS detector, i.e. for $x=0$;
    \item     $Q^{max}_{SEMSANS}(\delta_{SE})$ from eqs. \ref{eq:qmaxSEsetup};
    \item      $\xi_{exp}(\delta_{SE})$ from eq. \ref{eq:defineksiexp};

    \item    $G_{exp}(\delta_{SE})$ from eq. \ref{eq:GZfromIQhankelexp};

    \item $V_s(\delta_{SE})$    from eq. \ref{eq:newapp2} (via eq. \ref{eq:ss3}, eq. \ref{eq:ss4} and  eq. \ref{eq:ss5}, see Appendix C for details);
    \item  $T_{SANS}(\delta_{SE})$  from eq. \ref{eq:transscat} and $T_{SEMSANS}(\delta_{SE})$  from  eq. \ref{eq:tsemsans}.
\end{enumerate}

\section{\label{sec:specinficexamples}   Examples of applications }

\subsection{\label{subsec:raspberry}  Raspberry particles}
We first consider the example of raspberry particles 
which are present e.g. in Pickering emulsions  consisting of two immiscible phases, typically oil and
water. Small solid particles located at the oil-water interfaces
form an elastic "shell" that prevents coalescence.
Raspberry particles 
formed by adsorption of polystyrene latex particles on polydisperse oil droplets, schematically shown in Fig. \ref{fig:raspberry1}A,  have been  measured with SANS (down to $Q$ of 10\textsuperscript{-3}{} \AA\textsuperscript{-1}) and USANS (down to $Q$ of  8$\times$10\textsuperscript{-5}{} \AA\textsuperscript{-1}) in a combined study  \cite{Larson-SmithJacksonPozzo2012Raspberry}.  
 Maximum observable distances correspond to  0.63 $\mu$m and 7.8 $\mu$m, for SANS and USANS, respectively.
 The example of  experimental  and fitted $d\Sigma(Q)/d \Omega$ is shown in Fig. \ref{fig:raspberry1}B. 

The SEMSANS results for the maximum spin-echo length of 10 $\mu$m  and several sample thicknesses are shown in  Fig. \ref{fig:raspberrypolz1}.  
First of all, the overlap between SANS and SEMSANS is only achieved in the wide $Q$ mode of SKADI (Fig. \ref{fig:raspberrypolz1}B,D), because in this case 
the $\delta_{SE}$-region is broader due to the broader $\lambda$-range. 
Second, measurements at a longer sample-detector distance ($d_1$) lead to a lower $V_s(\delta_{SE})$ and are better suited for a more accurate SEMSANS modeling 
(compare Fig. \ref{fig:raspberrypolz1}B vs. Fig. \ref{fig:raspberrypolz1} D). The  lower $V(\delta_{SE})$ values can be explained as follows:
for larger $d_1$, $Q^{max}_{SEMSANS}$ gets smaller, so does $\Sigma_{exp}$ (cf. eq. \ref{eq:defineksiexp}), and, as a consequence, $V_s(\delta_{SE})$ approaches unity (cf.  eq. \ref{eq:pz}). 

Finally, the results for different sample thicknesses in Fig. \ref{fig:raspberrypolz1}B show that for larger thicknesses
  $V_s(\delta_{SE})$  becomes lower,  thus more appropriate for SEMSANS measurement. However, at the same time $T_{SANS}$ decreases, which implies higher multiple scattering effects on the SANS signal.   In all cases $T_{SEMSANS} > 0.9$, the multiple scattering impact on SEMSANS  is negligible.

\begin{figure}[htp]
\vcenteredhbox{ \begin{overpic}[scale=.3]{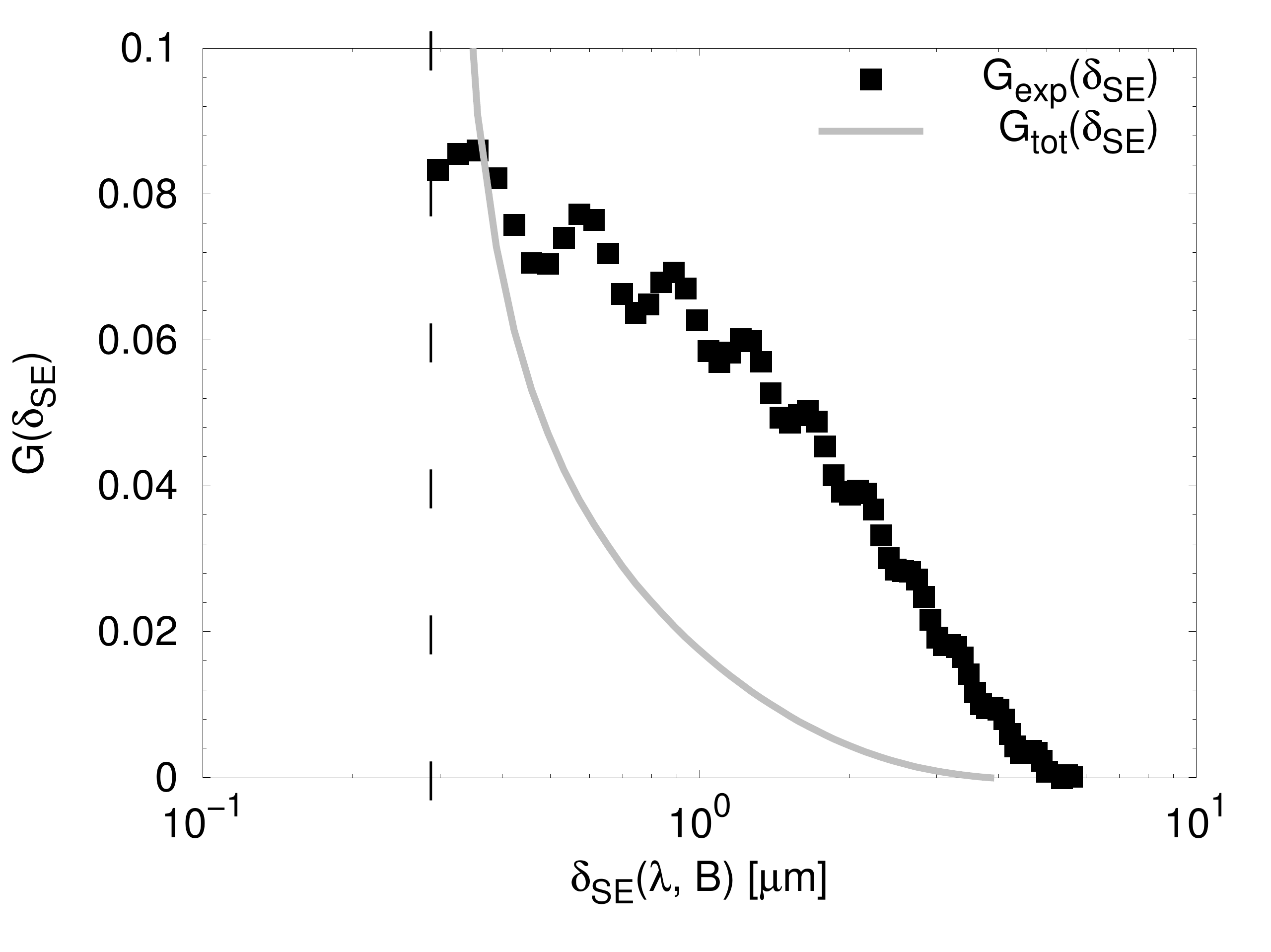}
\end{overpic} }
    \caption{  $G(\delta_{SE})$ for raspberry particles calculated from $d\Sigma(Q)/d \Omega$  via eq. \ref{eq:GZfromIQhankelexp} ($G_{exp}(\delta_{SE})$) and eq. \ref{eq:GZfromIQhankel} ($G_{tot}(\delta_{SE})$) for the wide $Q$ mode  and $\lambda_{min}=$3 \AA{} at $d_1=8$ m (the same conditions as in  Fig. \ref{fig:raspberrypolz1}B).
    A dashed vertical line corresponds to $\delta_{SE}=2\pi / Q^{min}_{SANS}$.}
\label{fig:raspberryGZ}
\end{figure}

Fig. \ref{fig:raspberryGZ} depicts the correlation functions $G_{exp}(\delta_{SE})$ and $G_{tot}(\delta_{SE})$ thus illustrating the effect of  the   finite upper integration limit  in eq. \ref{eq:GZfromIQhankelexp},  $Q^{max}_{SEMSANS}(\delta_{SE})$ due to the finite acceptance angle of the SEMSANS detector. 
$G(\delta_{SE})$ (cf.   eq. \ref{eq:GZfromIQhankel}) is a real space correlation function  \cite{AnderssonBouwman2008JAC}. Thus, the experimentally obtained $G_{exp}(\delta_{SE})$ is directly related to the structure of the sample. 
For example, in Fig.  \ref{fig:raspberryGZ},  $G_{exp}(\delta_{SE})$ becomes zero for $\delta_{SE}$, which exceeds the maximum correlation length of the structure. 
 Therefore, the latter can be estimated from the intercept of $G_{exp}(\delta_{SE})$ with the $\delta_{SE}$-axis. In fact this correlation length is larger than the  average diameter of a raspberry particle due to  the effect of polydispersity.  Note that the noticeable difference  between  $G_{exp}(\delta_{SE})$ and $G_{tot}(\delta_{SE})$  renders a visual analysis of $G_{exp}(\delta_{SE})$ less reliable and in some cases even impossible. However,  a proper, detailed analysis of  $G_{exp}(\delta_{SE})$   includes fitting of a Hankel transform of a structural model in terms of $d\Sigma(Q)/d \Omega$, and this calculation  does include the finite acceptance angle effect. 

\begin{figure}[tph]
\centering
\vcenteredhbox{
\begin{overpic}[scale=.40]{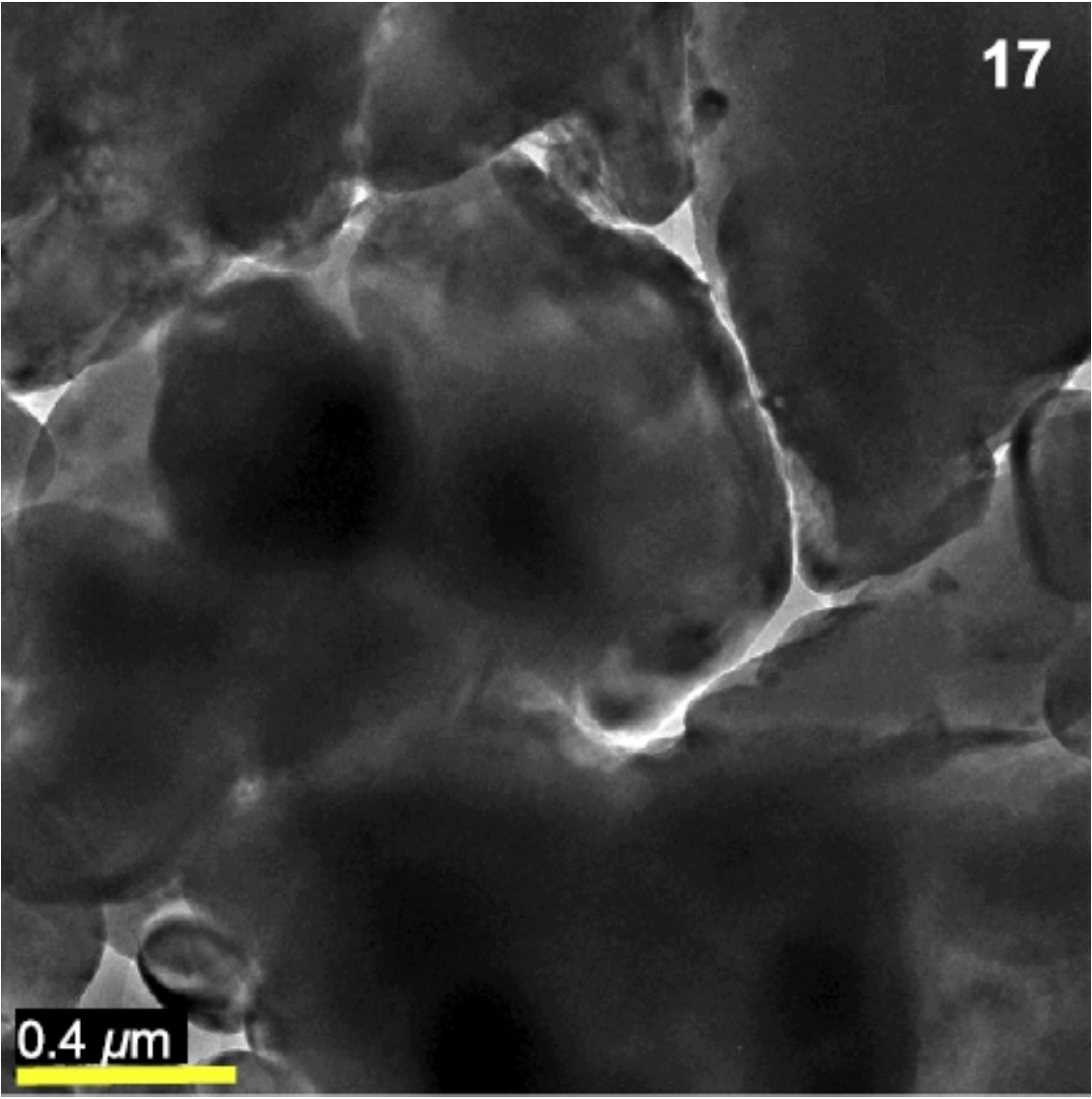}
 \put(-15,55){\bf{(A)}}
\end{overpic}
}
\vcenteredhbox{
\begin{overpic}[scale=.32]{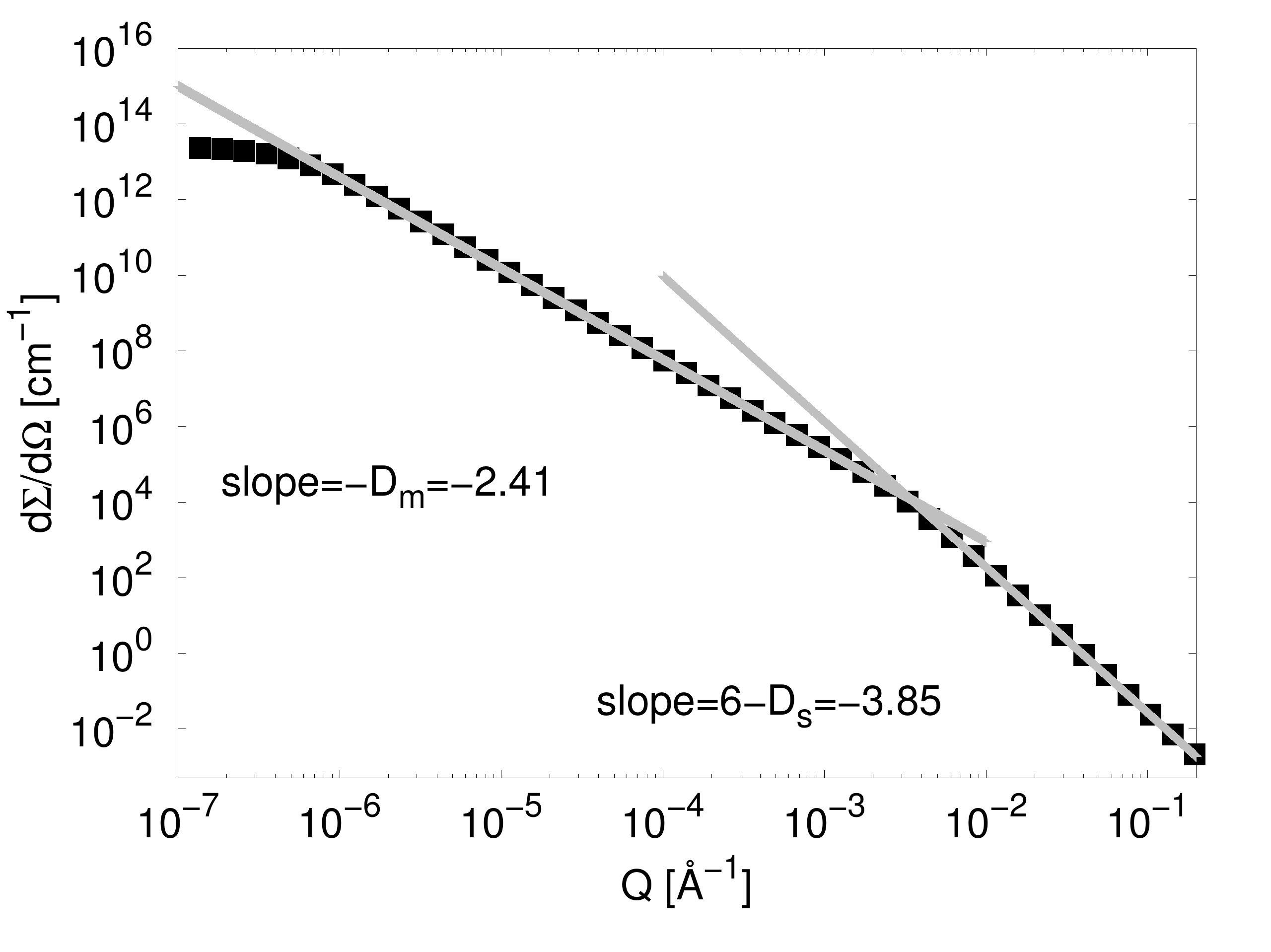}
    \put(50,60){\bf{(B)}}
\end{overpic}
}
\caption{A: a TEM image of chalk; reprinted with permission 
    from ref.  \cite{WangRother2013USANSIMAGEporous}, copyright (2013) Elsevier. B:  $d\Sigma(Q)/d \Omega$  of a chalk sample 
calculated  using the results  from a combined SANS, USANS and electron imaging study \cite{WangRother2013USANSIMAGEporous}.}
\label{fig:rocksIQ}
\end{figure}

\begin{figure*}[ht]
\centering
\vcenteredhbox{ \begin{overpic}[scale=.34]{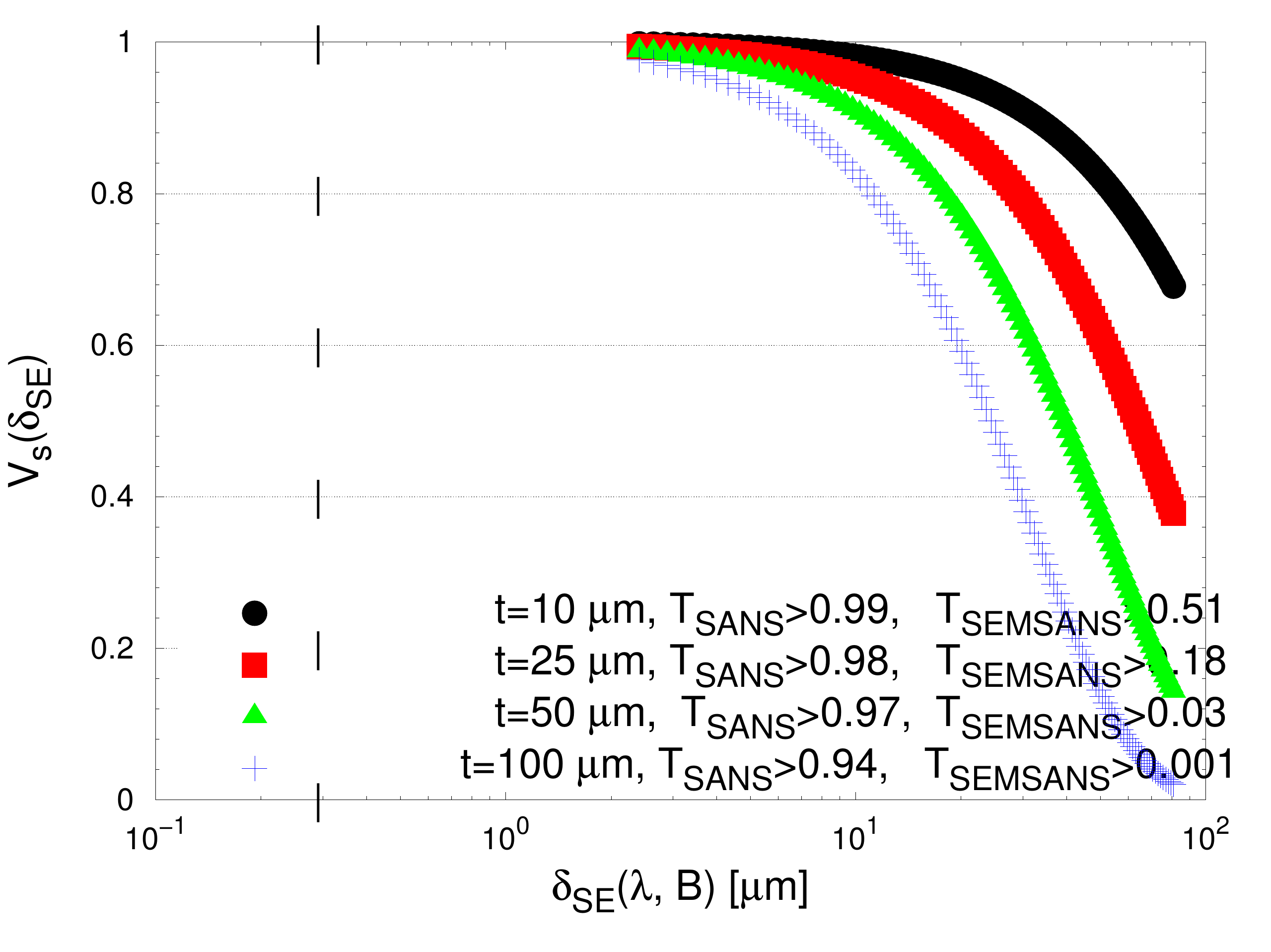}
        \put(60,36){\bf{(A)}}
\begin{picture}(0,0)
\put(7.5,30){\includegraphics[height=3.5cm]{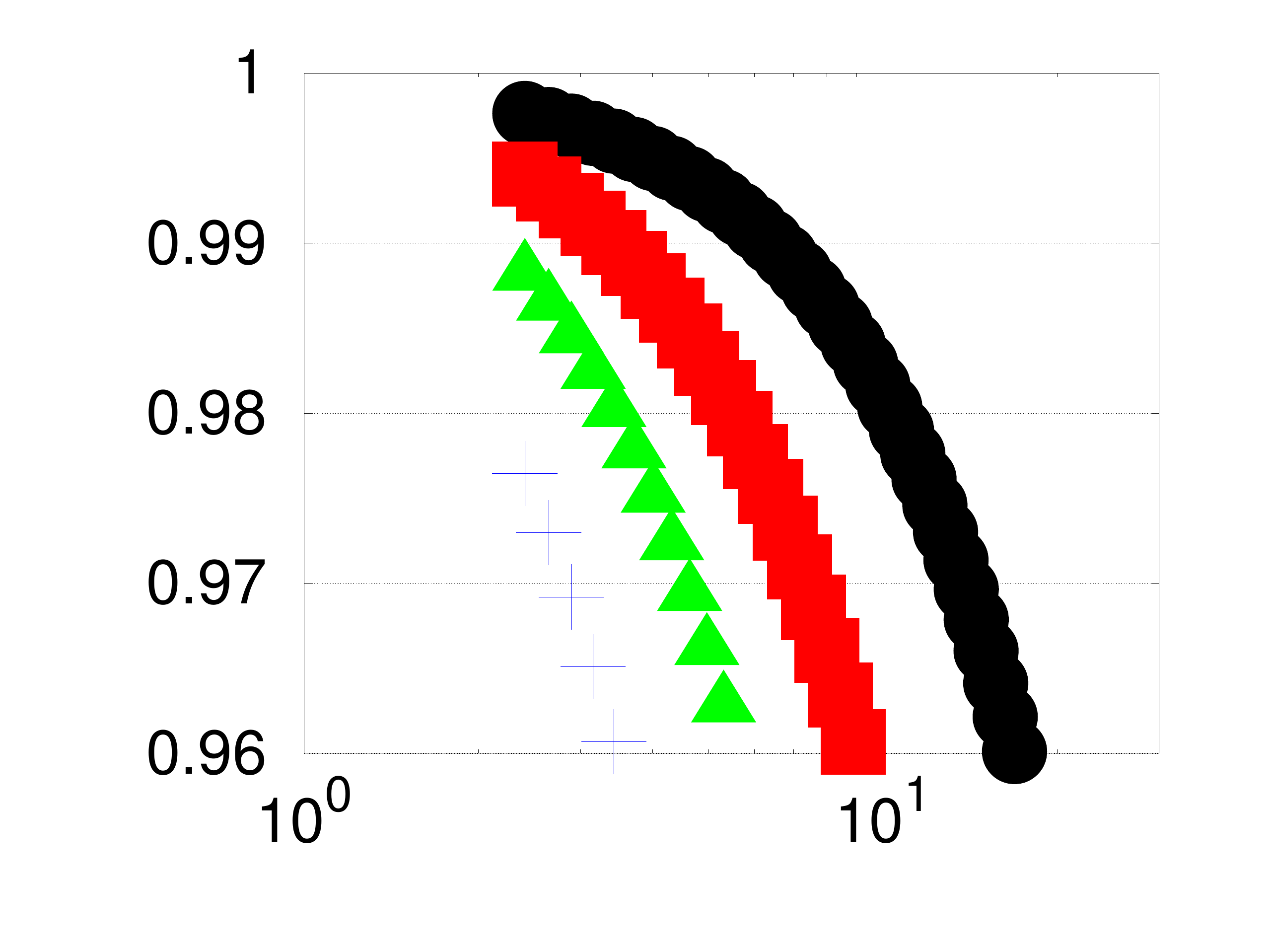}}
\end{picture}
\end{overpic} }
\vcenteredhbox{ \begin{overpic}[scale=.34]{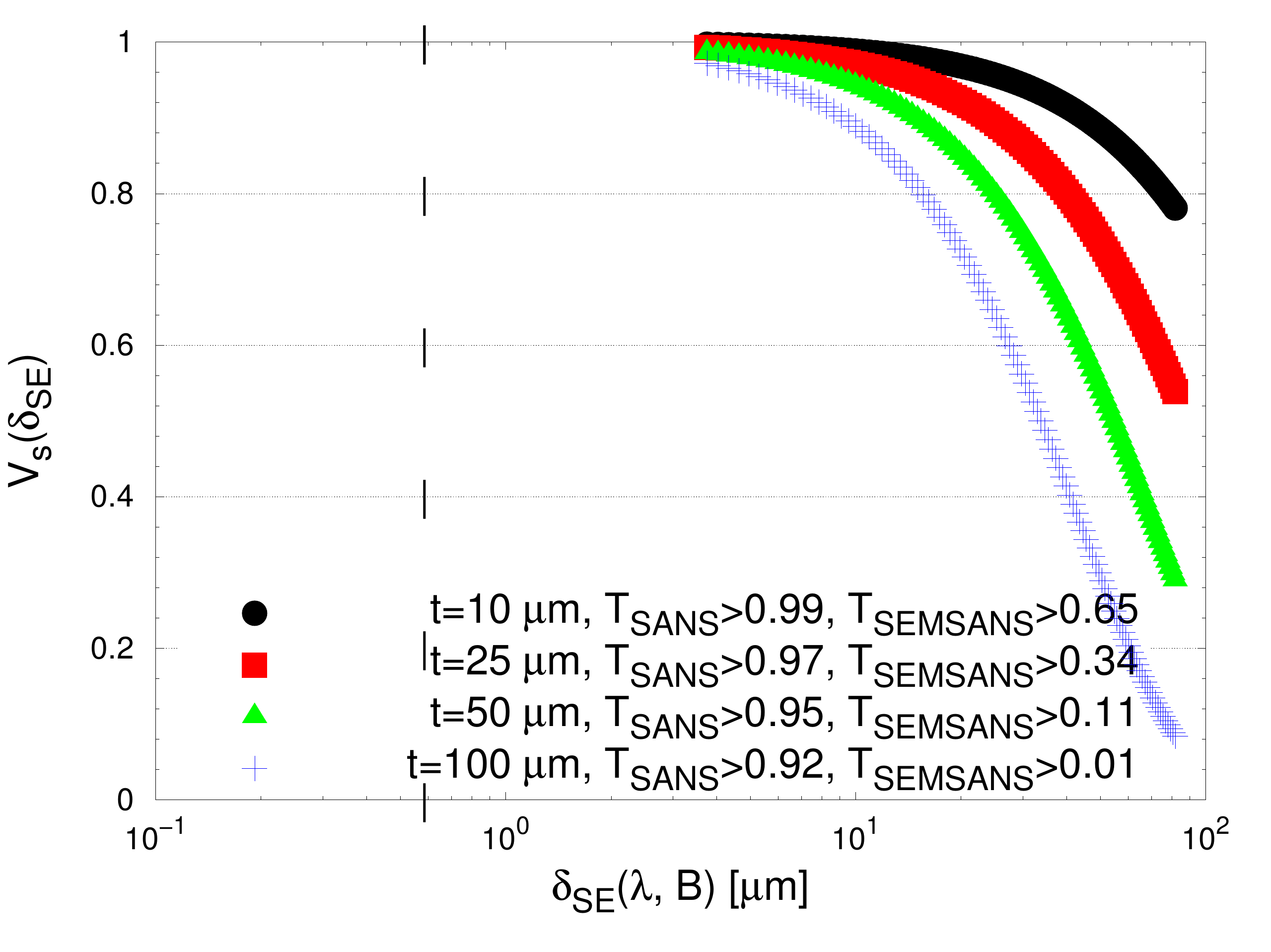}
        \put(60,36){\bf{(B)}}
\begin{picture}(0,0)
\put(7.5,30){\includegraphics[height=3.5cm]{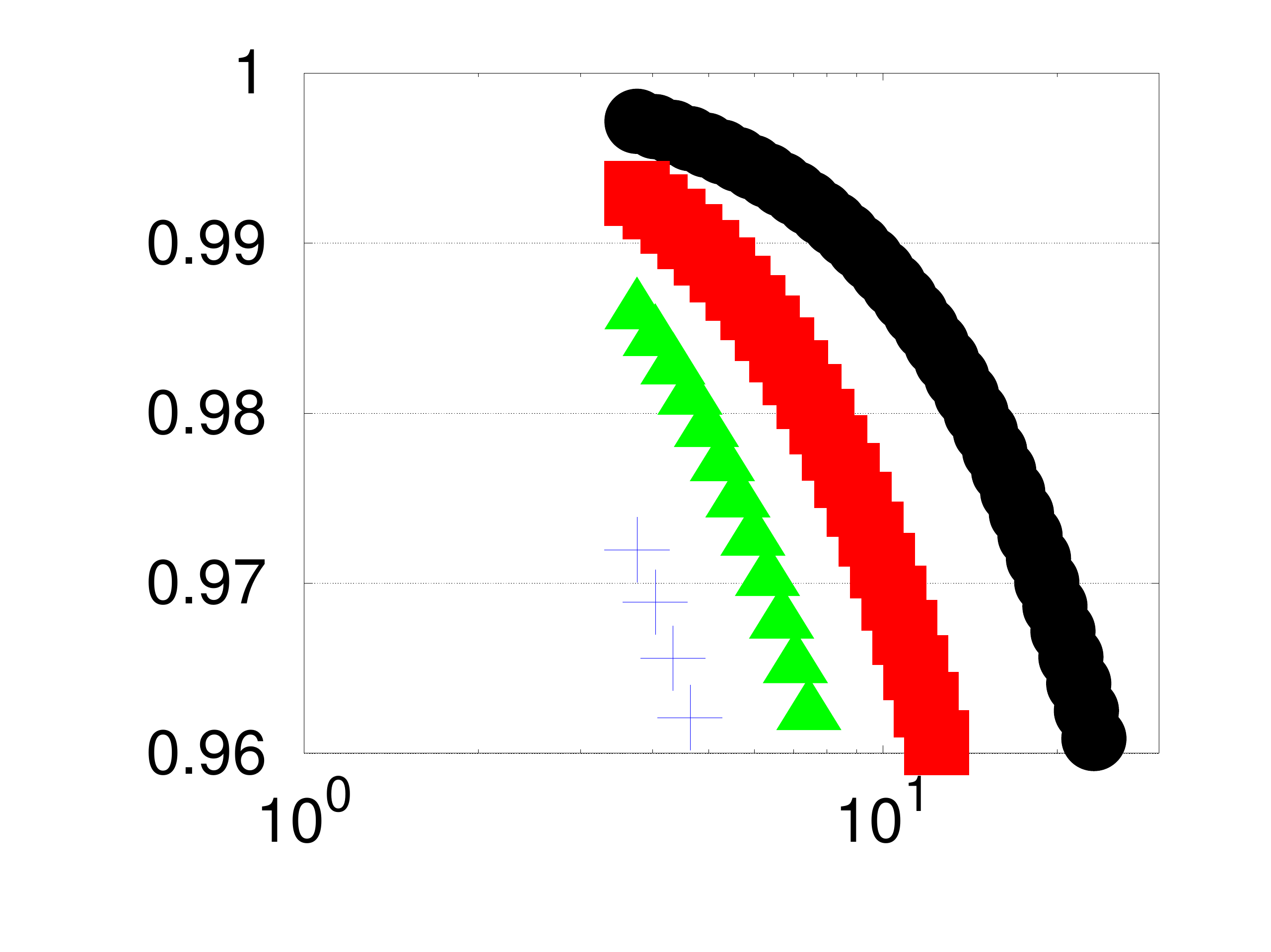}}
\end{picture}
\end{overpic} }
\caption{$V_s(\delta_{SE})$ for a chalk sample of several thicknesses ($t$), wide $Q$ mode,  $\lambda_{min} $= 3 \AA{}.
 A:  $d_1$=8 m,  $\lambda_{max}=$ 17.4 \AA, $B_1$=6.7 mT, $\zeta^{max}=$1 mm;  B:  $d_1$=20 m,  $\lambda_{max}=$ 14 \AA, $B_1$=9.8 mT, $\zeta^{max}=$1.6 mm. 
    Inset figures show that for the two thickest samples the condition $V_s(\delta_{SE})<0.99$ is met.
}
\label{fig:rockspolz1}
\vcenteredhbox{ \begin{overpic}[scale=.34]{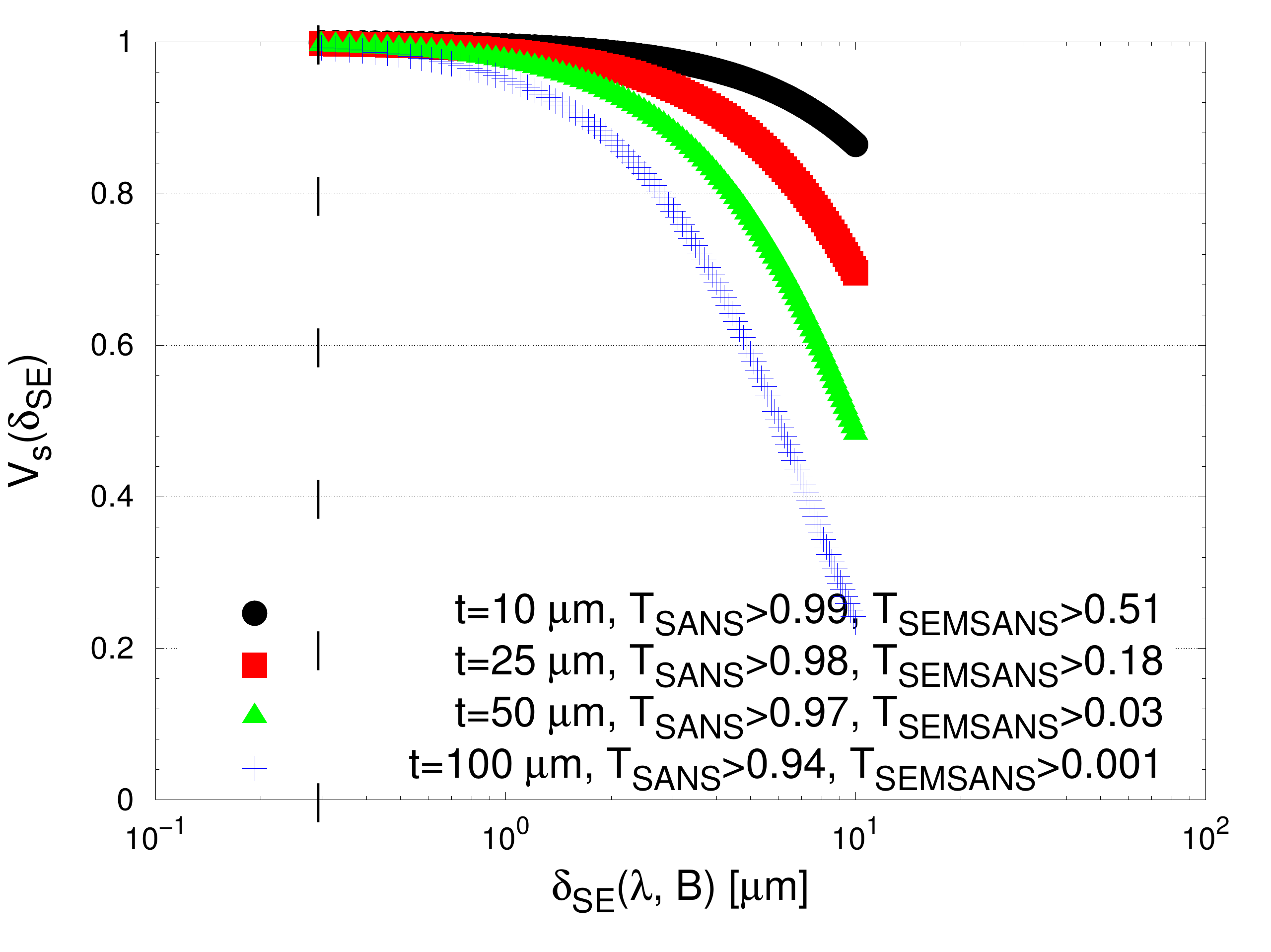}
        \put(75,36){\bf{(A)}}
\begin{picture}(0,0)
\put(7.5,25){\includegraphics[height=3.5cm]{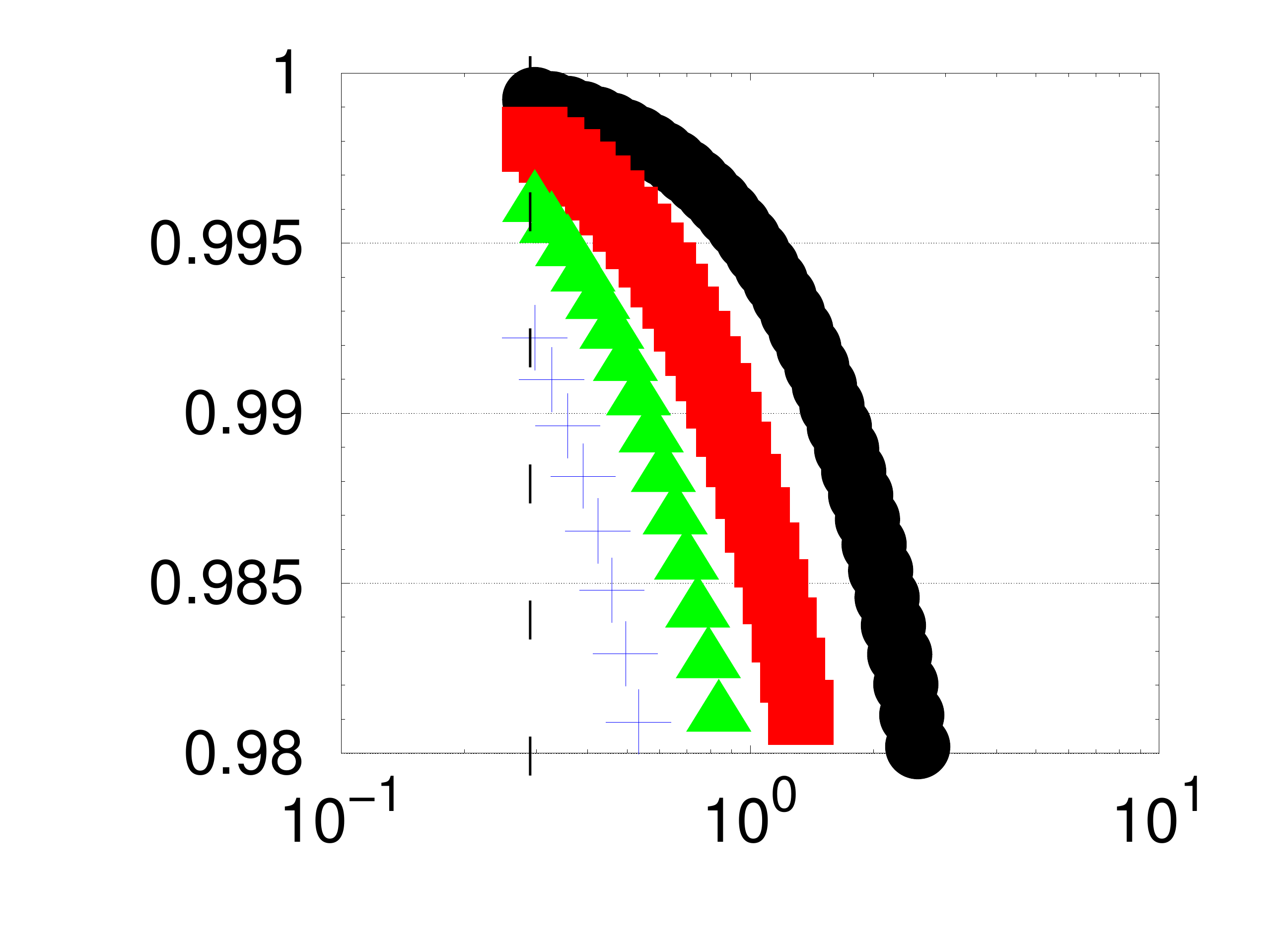}}
\end{picture}
\end{overpic} }
\vcenteredhbox{ \begin{overpic}[scale=.34]{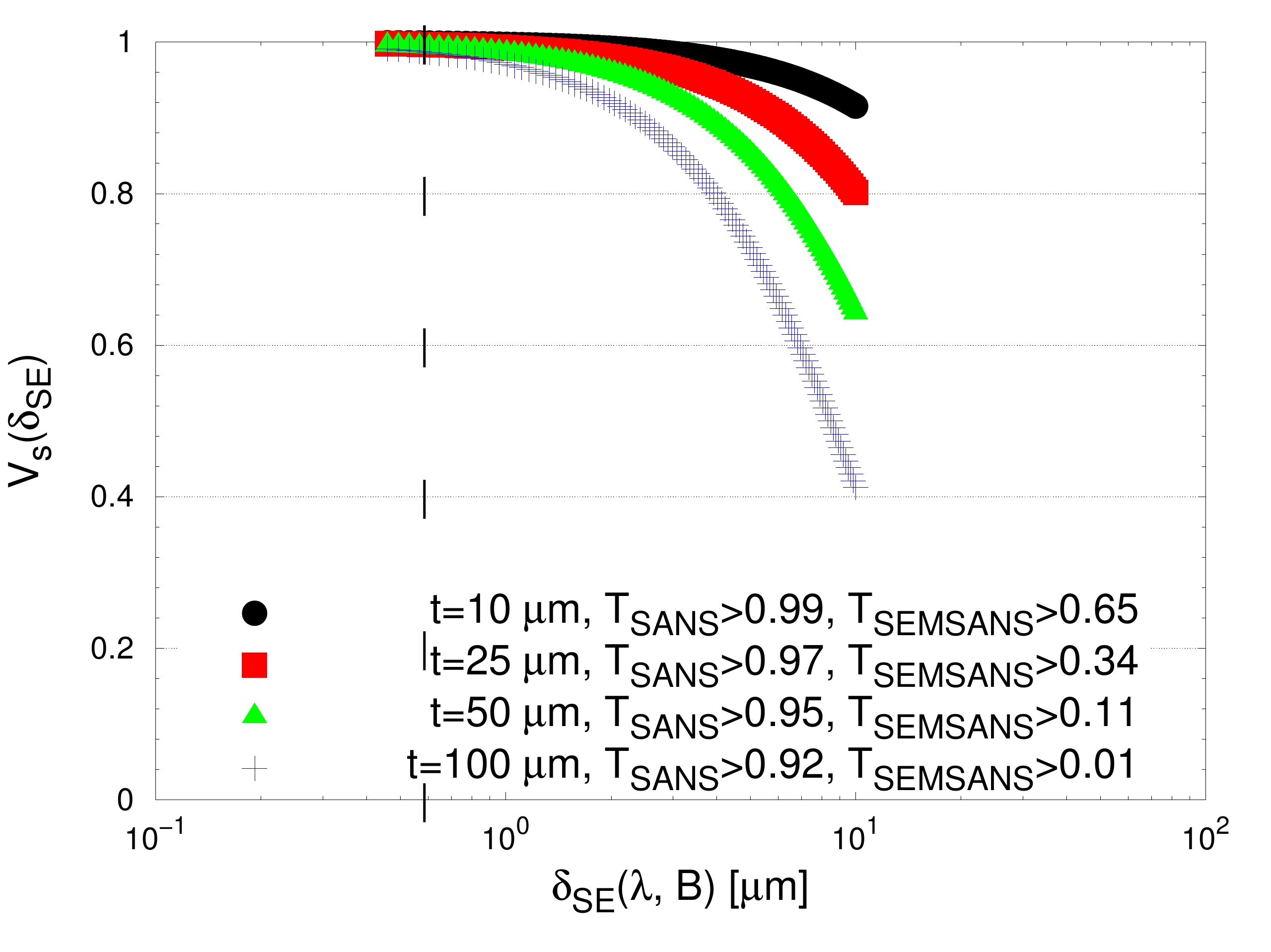}
        \put(75,36){\bf{(B)}}
\begin{picture}(0,0)
\put(7.5,25){\includegraphics[height=3.5cm]{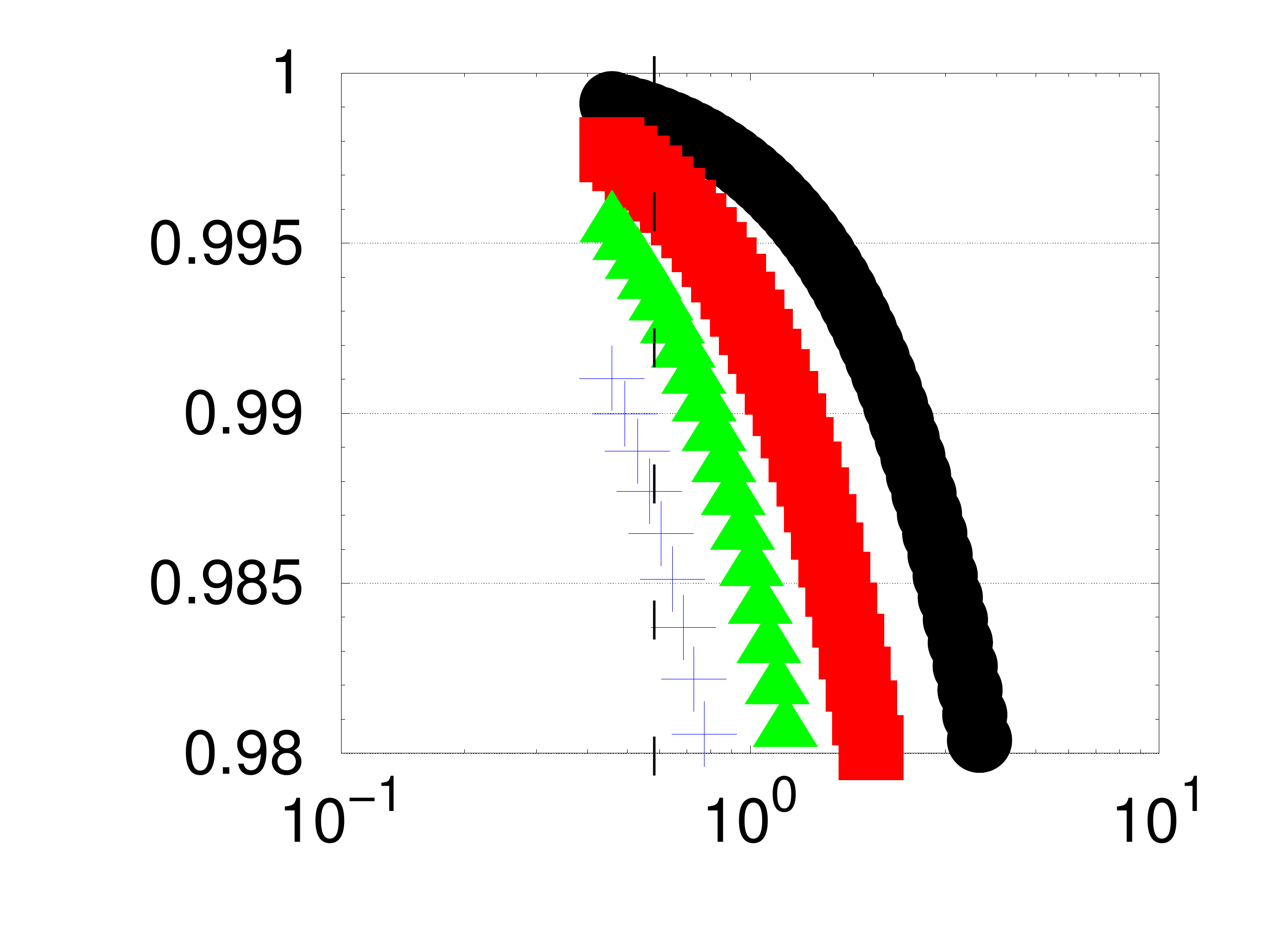}}
\end{picture}
\end{overpic} }
\caption{ $V_s(\delta_{SE})$ for a chalk sample of several thicknesses ($t$), wide $Q$ mode,  $\lambda_{min} $= 3 \AA{}, and for lower magnetic field settings than in  Fig. \ref{fig:rockspolz1}.
 A: $d_1$=8 m,  $\lambda_{max}=$ 17.4 \AA, $B_1$=2.4 mT, $\zeta^{max}=$8 mm;  B: $d_1$=20 m,  $\lambda_{max}=$ 14 \AA,  $B_1$=3.2 mT, $\zeta^{max}=$13 mm. Inset figures show that the condition $V_s(\delta_{SE})<0.99$ is met only for 
    the thickest sample in subfigure B  and only for $\delta_{SE}> 2\pi/Q^{min}_{SANS}$.
}
\label{fig:rockspolz2}
\end{figure*}
\subsection{\label{subsec:rocks} Fuel, rocks and minerals}
 SANS,  USANS  or their combination are commonly used  for structural characterizations   
 over  a wide range of length scales in geology, mineralogy or hydrocarbon recovery. 
 Examples of such studies are provided in diagenesis (change of a sedimentary rock into a different rock)  \cite{WangRother2013USANSIMAGEporous,AnovitzLittrell2013USANSporous}, pore structure in  coals  \cite{RadlinskiThiyagarajan2004RocksSANSCoal} and in nuclear graphite \cite{ZhouBouwmanPappas2016SESANSCARBON}, hydrocarbon generation  \cite{RadlinskiHope2000RocksSANSHydrocarbon}, tight gas reservoirs  \cite{ClarksonBlach2012RocksSANSGas} or gas shales   \cite{KingHuynh2015RocksSANSGasShale}. 
 These results are often complemented by other studies, which extend the structural information over length scales larger than 1 $\mu$m, like neutron imaging  \cite{ZhouBouwmanPappas2016SESANSCARBON} or transmission electron microscopy (TEM), such as the image of chalk in Fig. \ref{fig:rocksIQ}A.  
 
The differential cross-section of chalk  calculated from the  parameters obtained  from  SANS, USANS and imaging  is given in Fig. \ref{fig:rocksIQ}B. The slope of the curve in the $Q$-region covered by USANS and imaging gives information on the mass fractal dimension of the pore network ($D_m$).  The slope of the curve in the region of larger  $Q$-values (the region covered by SANS) gives the surface fractal dimension $D_s$ of individual pores. 

The original USANS experiment on chalk  \cite{WangRother2013USANSIMAGEporous} was performed  with an incident wavelength of $\lambda=2.38$ \AA{} and on a 0.15 mm thick sample in order to reduce multiple scattering effects.  For longer wavelengths such as those considered in this comparative study, the samples should be much thinner. 

The  $V_s(\delta_{SE})$ calculated for the chalk sample and for a maximum $\delta_{SE}$ of 100 $\mu$m is shown in Fig. \ref{fig:rockspolz1}. In this case, as it is important to cover a broad $\delta_{SE}$-range, only the wide  $Q$ mode is considered.  In addition the  $\delta_{SE}^{max}$ in a real experiment may not reach 100 $\mu$m because of limitations on the maximum magnetic field. In this configuration, a broad range of length scales is covered  but the $\delta_{SE}$-range has no overlap with the region accessed by SANS. In order to achieve an overlap between SANS and SEMSANS,  the SEMSANS measurements should be performed at a smaller magnetic field, such as in 
 Fig.  \ref{fig:rockspolz2}, where, however, $\delta_{SE}$ does not exceed $\approx$ 10 $\mu$m. Thus, the  technical constraints do not allow to reach an overlap between SANS and SEMSANS and at the same time  a maximum $\delta_{SE}$ of 100 $\mu$m.
  
 Another limitation occurs from multiple scattering effects. As $T_{SANS}>0.90$, multiple scattering does not significantly influence the SANS results. On the other hand, it does  limit the SEMSANS measurement because the  $T_{SEMSANS}$-values decrease to zero at spin echo lengths shorter than $\delta_{SE}^{max}$ for the largest sample thicknesses. A compromise between SANS and SEMSANS along the lines of Subsec. \ref{subsec:quality} leads to a maximum sample thickness of $\approx$ 50 $\mu$m, which, in some cases,  can not be easily achieved without modifying the mesoscopic structure of the sample. An interesting aspect of the small-angle scattering though is that it probes the structure  in the direction perpendicular to the incident beam, 
and thus SEMSANS can probe the structure even on the length scales larger than the sample thickness.

\begin{figure}[h]
\vcenteredhbox{
\centering
    \begin{overpic}[scale=.35]{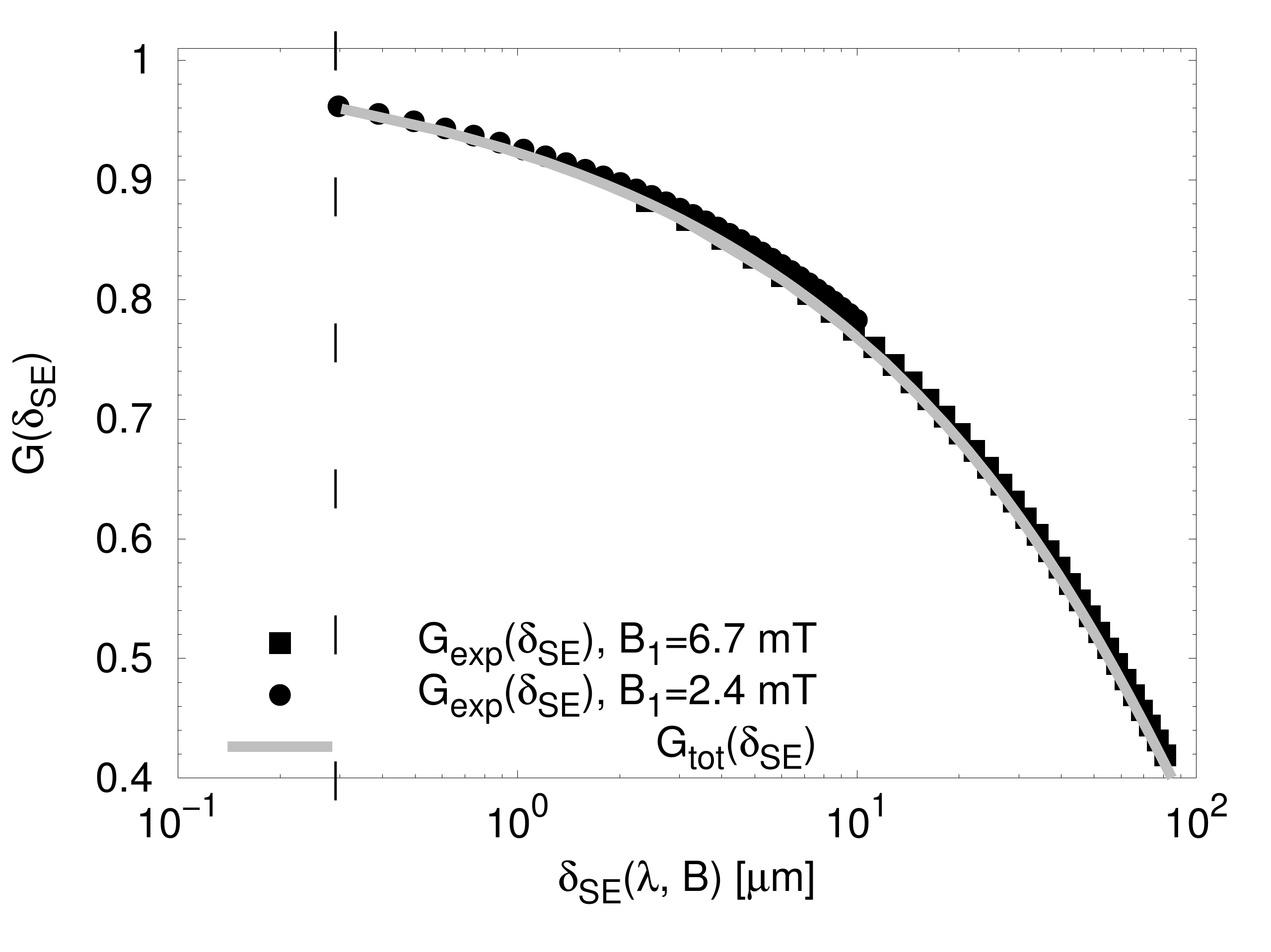}
\end{overpic} }
\caption{
    $G_{exp}(\delta_{SE})$ and $G_{tot}(\delta_{SE})$ for chalk sample, for the wide $Q$ mode, $d_1$=8 m, and $\lambda_{min}=$ 3 \AA{} (the total wavelength range is 3 \AA $< \lambda <$ 17.4 \AA).  
}\label{fig:rocksGZ1}
\end{figure}%

Calculated $G_{exp}(\delta_{SE})$ and $G_{tot}(\delta_{SE})$  are shown in Fig. \ref{fig:rocksGZ1}.
The difference between the two is rather small in contrast to the raspberry particle case given in Fig. \ref{fig:raspberryGZ}.
This small difference is due to the fact that the differential SANS cross-section of chalk decreases rapidly with increasing $Q$ and thus the cut-off at the high $Q$ region does not significantly alter the result of Hankel transform. 

%
%
%
%
%
\subsection{\label{subsec:stober} Growth of monodisperse spherical particles}
Monodisperse colloidal particles with sizes up to $\approx$1-2 $\mu$m have numerous applications \cite{XiaGatesLu2000MonoColloidSpheres}. 
The growth of such particles or kinetic processes involving their functionalization are potential science cases for a  combined SANS and SEMSANS instrument.
An example of such particles are "St\"ober particles", which  are monodisperse spherical silica particles  prepared by the  method described in 1968 by St\"ober et al.   \cite{StoberBohn1968SilicaParticles}.
%
%
%
%
%
%
\begin{figure}[h]
\centering
\vcenteredhbox{
\begin{overpic}[scale=.18]{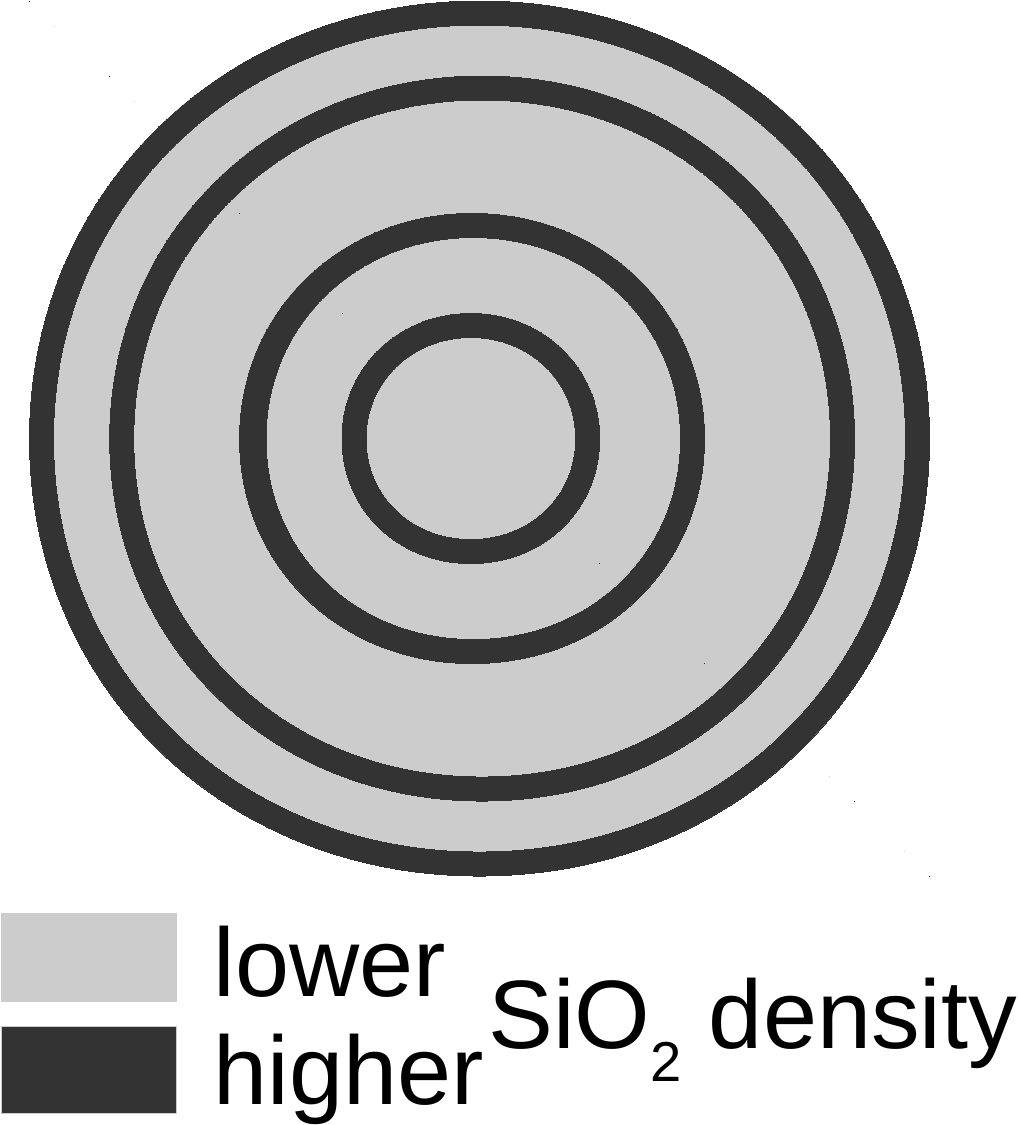}
    \put(50,105){\bf{(A)}}
\end{overpic}
}
\vcenteredhbox{
\begin{overpic}[scale=.24]{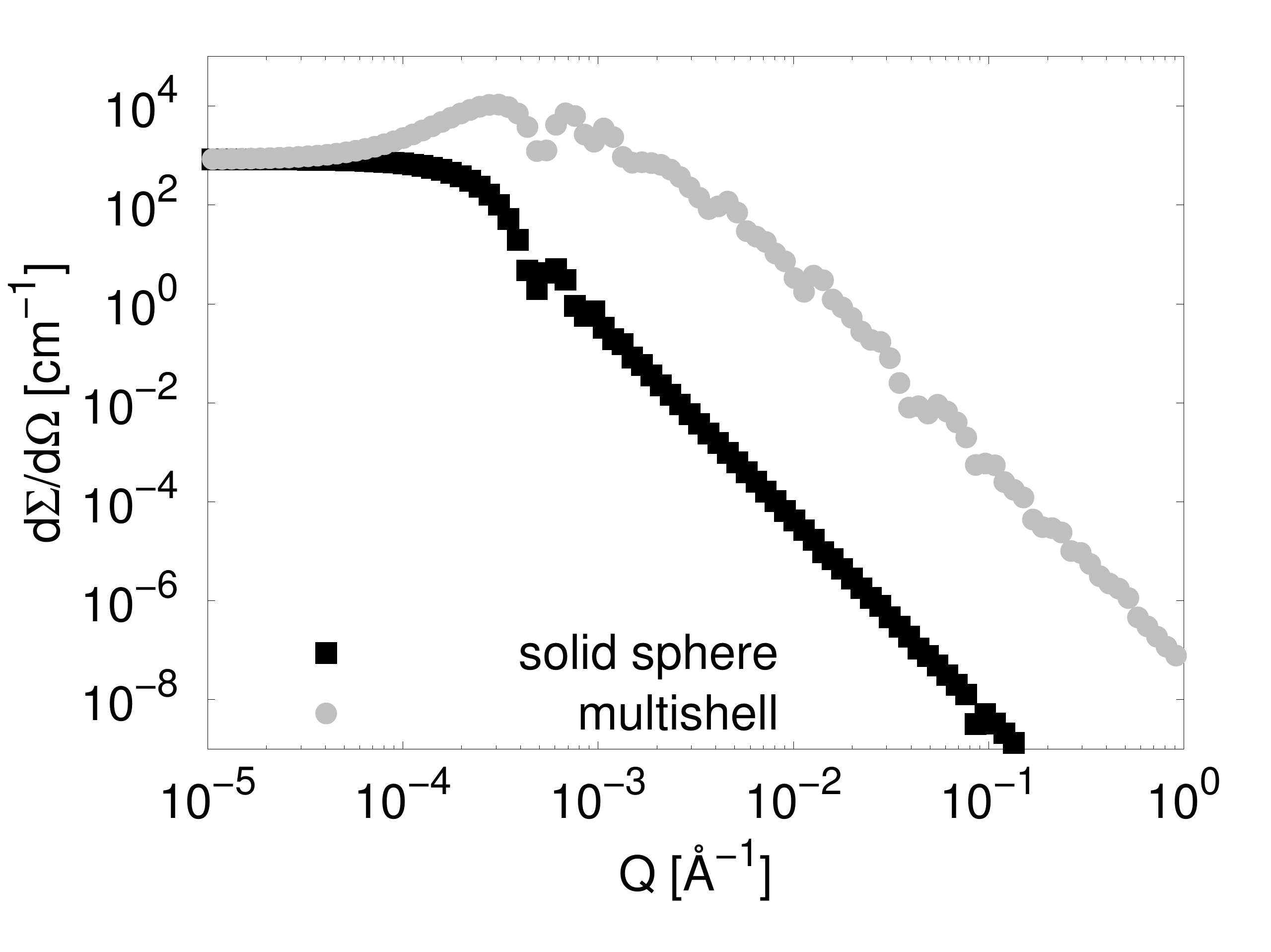}
    \put(80,50){\bf{(B)}}
\end{overpic}
}
    \caption{A: A model of  a \ce{SiO2} particle based on TEM images of large (diameter of $\approx$ 1.8 $\mu$m))  St\"ober particles prepared by a multistage method \cite{MasalovEmelchenko2013StoberSANS}. B: $d\Sigma(Q)/d \Omega$ for $d \lambda/\lambda=$  18 \% and in an aqueous solution (\ce{H2O}:\ce{D2O}$\approx$1:1), for two models of St\"ober particles: solid spheres with a radius of $\approx$ 1 $\mu m$, and  multishell spheres (spheres of the same size and average density, but with nine concentric shells which  have a higher density than the density between the shells).}
\label{fig:stoberTEM}
\end{figure}

A number of SAXS and USAXS experiments have been performed on  St\"ober  particles with diameters in the range from 10 to 100 nm, depending on  reaction times and conditions \cite{BoukariHarris2000StoberUSAXSTimeResolved,GutscheNirschl2014SAXSTimeResolvedStober}.
St\"ober particles prepared by a multistage method may have much larger diameters, larger than 
 1 $\mu$m \cite{MasalovEmelchenko2011StoberInternal}.
Transmission electron microscopy shows that such  St\"ober particles are not homogeneous but contain a series  of thin ($\approx$ 15 nm thick) concentric spherical shells of larger density  \cite{MasalovEmelchenko2013StoberSANS} as schematically shown in Fig. \ref{fig:stoberTEM}A.

USAXS experiments on   St\"ober particles  were performed with  $Q$ larger  
than $\approx2\times$10\textsuperscript{-3} \AA\textsuperscript{-1}{}, which corresponds  to a maximum  observable length of $\approx $ 300 nm. 
Longer length scales (up to 500 nm) have been studied recently with SESANS \cite{ParnellWashingtonPynn2016SESANS}. 
Time-resolved combined  SANS and SEMSANS experiments would allow to cover  a wide range of length scales and observe in-situ the growing  St\"ober particle 
as a  function of reaction time. This will provide important input to understand how particles with tailored properties  can be produced and 
  at the same time discriminate between various models of their inner structure. 

We have modeled St\"ober particles prepared by a multistage method  as homogeneous solid spheres with a radius of $\approx $ 1 $\mu$m and as spheres
 of the same radius and the same average density but with several thin concentric shells with increased density. Sphere radii and the concentric shells were taken from an earlier study  \cite{MasalovEmelchenko2013StoberSANS} and details are given in the supplement.
\begin{figure*}[t]
\centering
\vcenteredhbox{ \begin{overpic}[scale=.34]{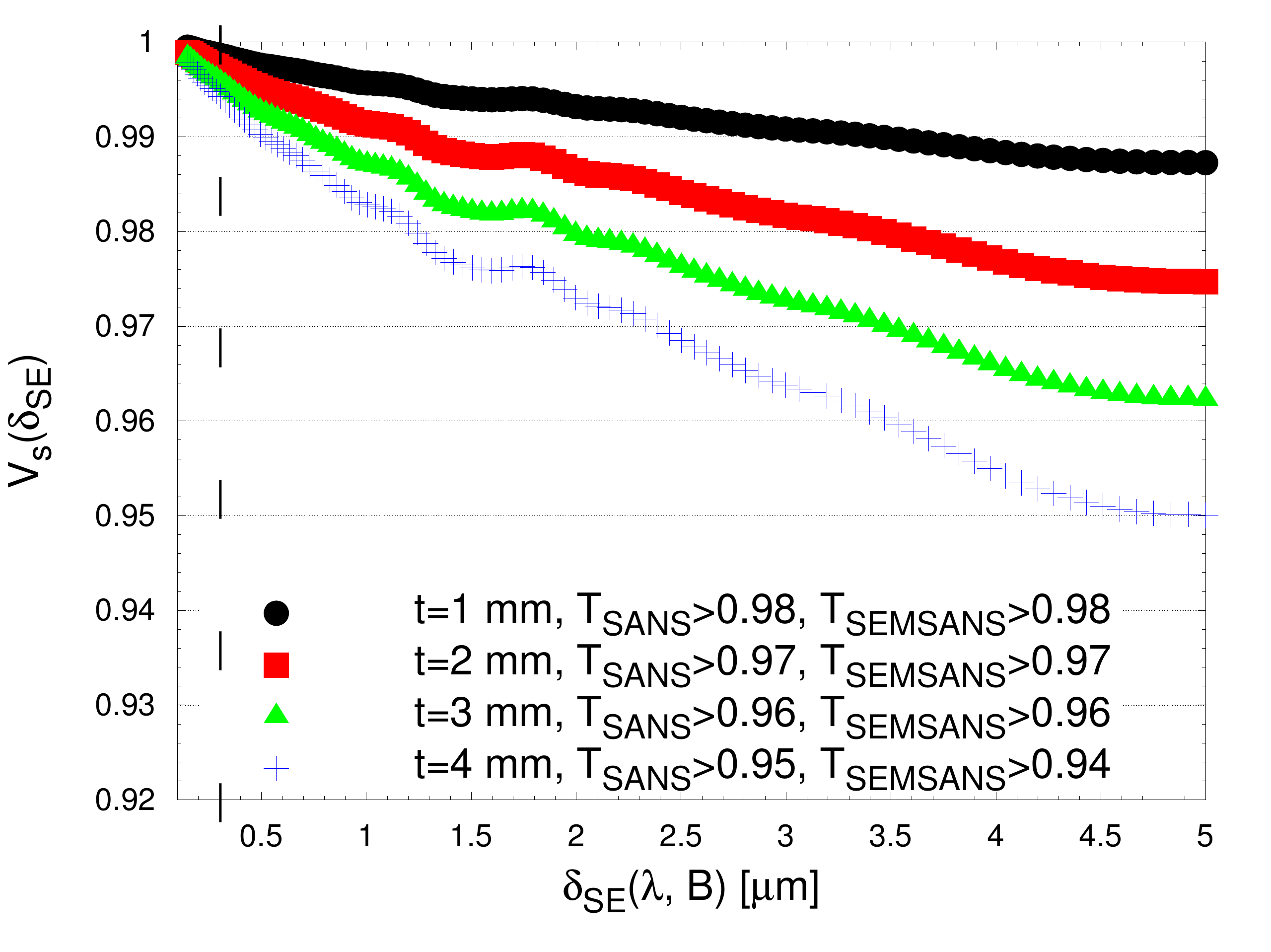}
        \put(25,36){\bf{(A)}}
\end{overpic} }
\vcenteredhbox{ \begin{overpic}[scale=.34]{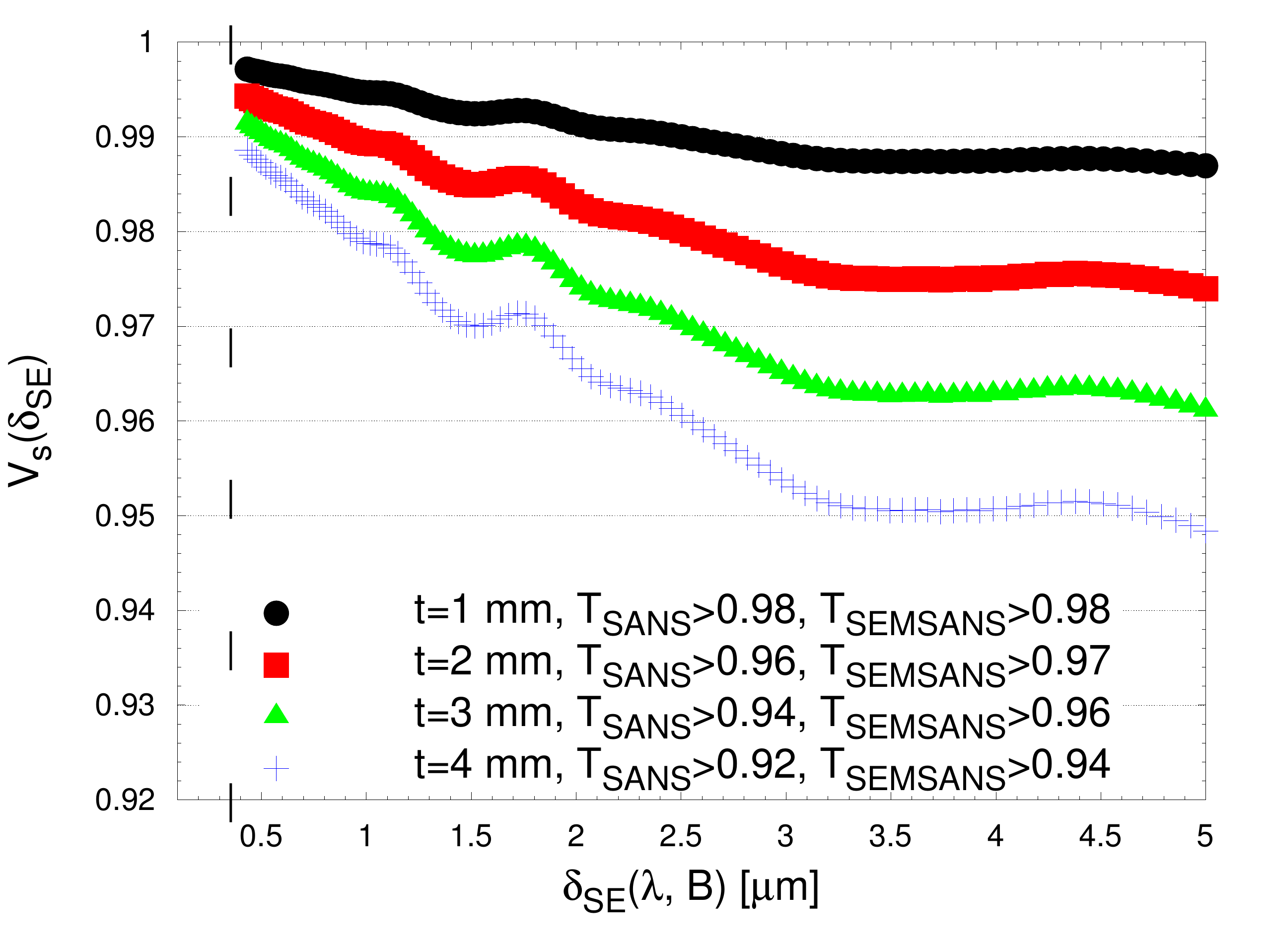}
        \put(25,36){\bf{(B)}}
\end{overpic} }
\caption{Experimental $V_s(\delta_{SE})$  for a sample of  St\"ober particles modeled by multishell spheres  
    of  several thicknesses ($t$), for wide $Q$ mode and $d_1$=8 m. A: $\lambda_{min}$=3 \AA{}, $\lambda_{max}$=17.4 \AA{},  $B_1$=0.4 mT, $\zeta^{max}=$16 mm;  B: $\lambda_{min}$=6 \AA{}, $\lambda_{max}$=20.4 \AA{}, $B_1$=0.3 mT, $\zeta^{max}=$11 mm.  A dashed vertical line corresponds to $\delta_{SE}=2\pi / Q^{min}_{SANS}$.  }
\label{fig:stoberPZ1}
\end{figure*}%

The resulting differential SANS cross-sections are shown in Fig. \ref{fig:stoberTEM}B  for an aqueous solution  (\ce{H2O}:\ce{D2O}$\approx$1:1).
     Approximate matching of the solvent scattering length density (SLD) to the average SLD of a multishell particle (the contrast is just $\approx 5\times 10^7$ cm\textsuperscript{-2}) allows to distinguish between  two different
models of internal structure. 

Fig. \ref{fig:stoberPZ1} provides examples of the calculated $V_s(\delta_{SE})$ for the case where a broad wavelength range is used and $d_1=8$ m. A larger $d_1$-distance would decrease the acceptance angle and lead to $V_s(\delta_{SE})$ values that are too close to unity, similarly to the raspberry particles case. Also, if $\lambda_{min}$ increases, the minimum $\delta_{SE}$ increases, and the overlap between SANS and SEMSANS shrinks.

 For all sample thicknesses and for both wavelength ranges,  $T_{SANS}$ and  $T_{SEMSANS}$ are always higher than 90 \%, indicating that multiple scattering is negligible for both SANS and SEMSANS.     
 The deduced minimum total sample transmission (determined from eq. \ref{eq:transtwoparts}) for $t=$ 4 mm is 30 \% at $\lambda^{min}$=3 \AA{} (when $\lambda^{max} =$ 17.4 \AA) and  24\% at $\lambda^{min}$=6 \AA{} (when $\lambda^{max} =$ 20.4 \AA). 
 
\begin{figure}[h]
\vcenteredhbox{ \begin{overpic}[scale=.35]{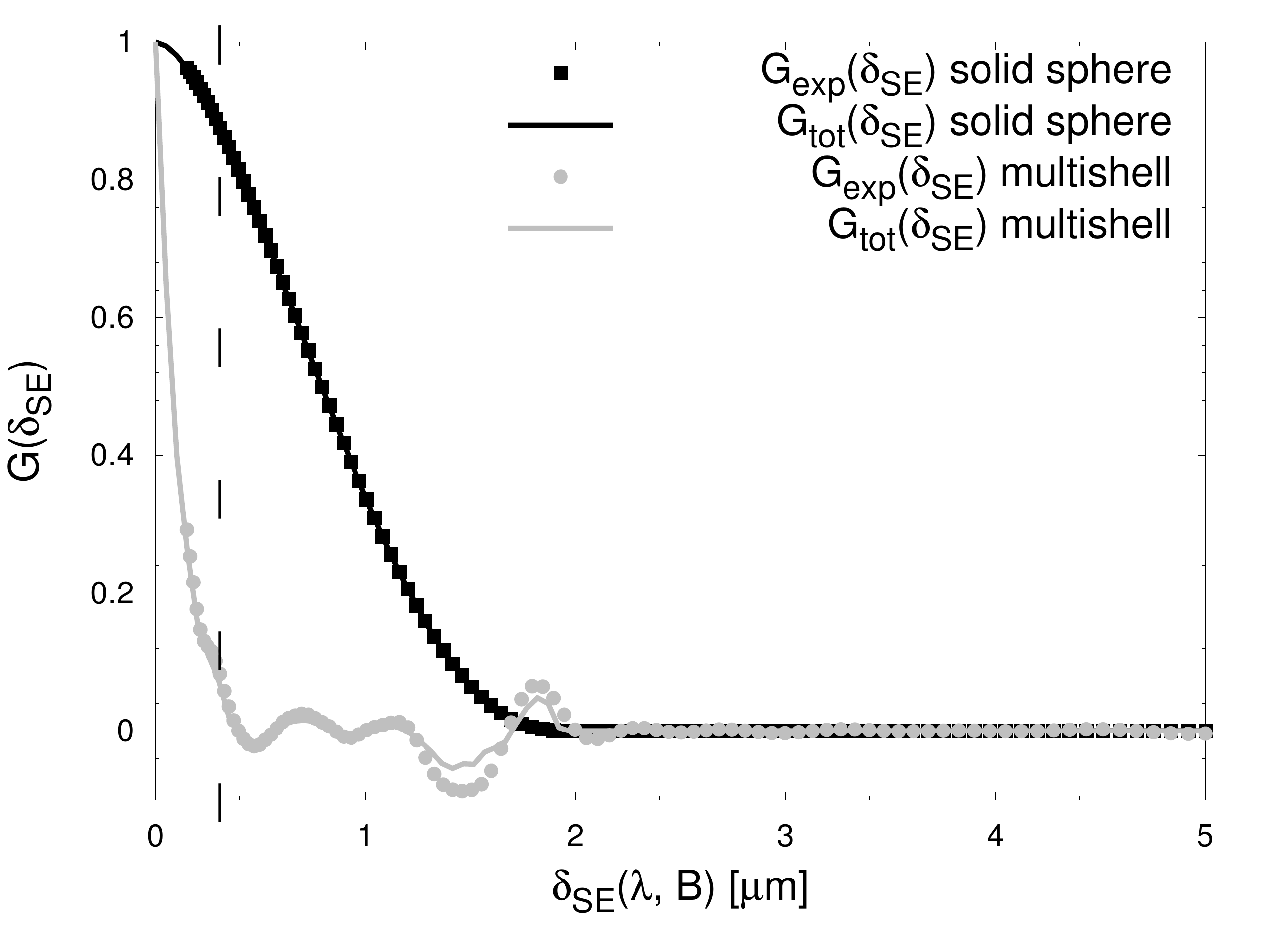}
\end{overpic} }
    \caption{ $G_{exp}(\delta_{SE})$  and $G_{tot}(\delta_{SE})$  for St\"ober particles modeled by solid and multishell spheres, for  $\lambda_{min}=$ 3 \AA{},  and other conditions corresponding to Fig. \ref{fig:stoberPZ1}A. The peaks in $G(\delta_{SE})$ correspond to diameters of the shells with increased densities (outer shell radii are given in the supplementary material).
}\label{fig:stoberGZ1}
\end{figure}

    An example of $G_{exp}(\delta_{SE})$  and  $G_{tot}(\delta_{SE})$ is shown in Fig. \ref{fig:stoberGZ1}.
  It can be seen that the solid spheres $G(\delta_{SE})$  differs from that of multishell spheres.  Thus,  it is  possible to distinguish  between 
  the two models using  SEMSANS.  In addition, $G_{tot}(\delta_{SE})$ and $G_{exp}(\delta_{SE})$ are close to each other for solid spheres
   but are quite different for multishell spheres.  This reflects the fact that for multishells, the ratio between the differential scattering cross-section  at large $Q$ and small $Q$ is larger than for solid spheres.  Therefore, the cut-off of the high $Q$ region via an upper integration limit  in eq. \ref{eq:GZfromIQhankelexp} has a larger effect on $G_{exp}(\delta_{SE})$ for multishell than for solid spheres.
\section{\label{sec:discussion}Discussion}

\subsection{\label{subsec:discussion1} Potential applications}
The three  examples presented above represent  three different science cases  illustrating potential scientific applications for a combined SANS and SEMSANS instrument.
In the case represented by the raspberry particles, the macroscopic incoherent scattering cross-section is low, 
and the scattering by large particles is significant but not very strong. It is thus possible to combine SANS and SEMSANS to cover a wide range of length scales in a single measurement.
In the case of a strong SANS scatterer like chalk, which scatters across a large range of length scales, it is not possible to cover the wide range of interest in a single measurement without a gap, which however, amounts to just a few micrometers out of the total range of about four orders of magnitude, from 1 nm to 80 $\mu$m.
In the last case represented by an aqueous solution of St\"ober particles, the macroscopic incoherent scattering cross-section is large but the coherent scattering contribution at small $Q$ is relatively weak.  Here, too,  a wide range of length scales can be covered in a single measurement  but the total transmission may be as low as 30 \% although with negligible multiple SANS contribution. For all three cases a sample thickness can be found that keeps the impact of multiple scattering at an acceptable level and at the same time leads to reasonable SEMSANS signals, $V_{SE} \le$ 0.99.

The specific parameters of the SEMSANS add-on considered here lead to a range of length scales covered without  a gap between 
the regions accessed by SANS and SEMSANS  extending up to $\approx$ 10 $\mu$m. 
A larger range can be covered but with a gap or by performing SEMSANS measurements with more than one magnetic field setting, but what are the implications?

The magnetic field settings do not interfere with the SANS measurements as long as both "spin-up" and "spin-down" intensities are recorded for every magnetic field. Therefore, with respect to SANS, the use of more than one magnetic field has just one small disadvantage: the  SANS measurement  time is reduced by the time required to change magnetic field, which is a few seconds.
For SEMSANS, the choice of several magnetic fields settings will only affect the time resolution (for kinetic studies), which can be pondered by the potential benefit of covering a much broader length range.

\subsection{\label{subsec:discussionfeasibility} Feasibility}

SANS is a well-established technique,  and the feasibility of SEMSANS experiments has also been proven \cite{BouwmanDuifGaehler2009Larmor,StrobBouwmanlPlompTOFSEMSANS2012,StroblWiederBouwman2012SEMSANS,SalesStrobl2015SEMSANS,StroblSalesHabicht2015SEMSANSNature}. Hence, the  combination of  SANS with SEMSANS depends on the specific scientific question and required measurement times. 


Section \ref{subsec:SANSeffect}  showed that the primary effect of a SEMSANS add-on on a SANS instrument  consists in the reduction of incident
beam intensity by a factor between 5 and 10, depending on the performance of the polariser and analyser in front of the sample. Furthermore, in order to make sure that the measured SANS cross-sections are not affected by the intensity modulation, the SANS intensities must be adequately averaged to obtain the shim intensity, which may decrease the time resolution as well, typically by a factor of two. Indeed, the maximum oscillation periods provided for all application examples, and the estimation of the reduction in the oscillation amplitude given in
Sec. \ref{subsec:SANSeffect}, indicate that such an averaging will be required in many cases. Thus the question is:  what is the time resolution that can be achieved for kinetic SANS measurements with adequate counting statistics? In general,  for time-resolved measurements the time resolutions of SANS and SEMSANS do not need to match each other.  Time binning can be done so as to ensure that   counting statistics is acceptable for both SANS and SEMSANS. Such an adaptive binning is especially easy in the case when  data acquisition is done in the "event mode".
  
  The examples of applications show that sample thicknesses suitable for  combined SANS and SEMSANS measurements are similar to those that would be chosen for a stand-alone SANS measurement, thus keeping multiple scattering effects to an adequate level. Consequently, the SEMSANS add-on would reduce the time resolution by a factor $10 \times 2=20$ with respect to a   stand-alone SANS measurement.
  This is of course a severe drawback, which however may be compensated by the high flux of the ESS. The incident neutron beams at the ESS are expected to be at least 10 times more intense than in other sources, which would render combined SANS and SEMSANS time resolved measurements an attractive option. These would be at least as good as SANS measurements nowadays, but with an additional structural information provided by SEMSANS.

Tentative measurement times and time resolutions for SEMSANS measurements can be estimated by considering that a large part of intensity measured by the SEMSANS detector comes from the transmitted neutron beam. Therefore, even if only a  narrow part in the center of MCP detector is used (e.g.  for $|x|<$ 1 mm, in order  to have a symmetric range of accepted scattering angles for all $x$, see eq. \ref{eq:thetamaxSEsetup}), the counting statistics should not  be a  problem. In addition, the high brilliance of the incident beam at ESS will not pose a problem for the SEMSANS detector. High counting rates reaching 10\textsuperscript{8} cm\textsuperscript{-2}s\textsuperscript{-1} can be handled by state-of-the-art MCP detectors, with at the same time a spatial resolution of 55 $\mu$m  \cite{TremsinFilges2012SEMSANSDETECTOR}.


\subsection{\label{subsec:discussion2}An effect of  finite acceptance angles}

The examples above show that for the raspberry particles and multishell St\"ober particles  $G_{exp}(\delta_{SE})$  is quite different from  $G_{tot}(\delta_{SE})$, while for rocks and large solid  St\"ober particles the difference between the two is  negligible.
This difference is caused by  the finite acceptance angle of the SEMSANS detector. In the cases where the differential scattering cross-section  is a fast decaying function of $Q$ (e.g. for fractal structures), the cut-off  at $Q$-values larger than  $Q^{max}_{SEMSANS}$ has a negligible effect on the Hankel transform.

However, as in general the  structure of a sample  is unknown, it is  impossible to predict the effect of acceptance angle. In this case a  
  visual analysis of $G_{exp}(\delta_{SE})$ may lead to incorrect conclusions and requires the fit of a SANS model on  $G_{exp}(\delta_{SE})$  using  eqs.  \ref{eq:defineksiexp}-\ref{eq:GZfromIQhankelexp} by taking into account the acceptance angle and the corresponding 
  $Q^{max}_{SEMSANS}$.   This is analogous to the  angular and wavelength resolution  effects in SANS, which affect the positions of  peaks or  minima  but can be taken into account during the fit.


%
\subsection{\label{subsec:strategymethods}  The choice of experimental parameters}
The major application for combining SANS and SEMSANS in a single measurements are time-resolved studies of kinetic processes 
and multiscale phenomena covering a large region of length scales. These are the cases where the substantial decrease in intensity caused by the SEMSANS add-on can be justified. 

On the other hand,  the largest range of spin-echo lengths and the best $V(\delta_{SE})$-signal can be achieved for just one  configuration of SKADI,
namely for a sample to detector distance of 8 m in the wide $Q$ mode. Because the neutron flux decreases dramatically for longer wavelengths, such a combined setup should use a minimum wavelength achievable for a polarised neutron beam (for SKADI, it is  $\lambda_{min}$ = 3 \AA). This minimum wavelength would also maximise  the overlap between the SEMSANS and SANS length scales. In such a case the choice of magnetic field depends on the maximum $\delta_{SE}$ that should be reached. 

Thus, the only remaining free parameter is the sample thickness, the  choice of which
will affect the multiple SANS effect on the measured differential SANS cross-sections, and possibly, 
on the  maximum spin-echo length that can be observed with SEMSANS.  
The sample thickness will also affect the SANS counting statistics data and the quality of the SEMSANS data (that is,  to what extent $V(\delta_{SE})$ differs from 1).
However,  multiple scattering depends on the structure of each specific sample and there is no simple recipe on how to select an optimal sample thickness. 
This must be  estimated  using   sample transmission, SEMSANS signal, and counting statistics obtained by test measurements or by calculations.

\subsection{\label{subsec:discussion3}Simultaneous analysis of SANS+SEMSANS results}
A simultaneous analysis of the SANS and SEMSANS data requires the transformation of either 1)  $G_{exp}(\delta_{SE})$  to $d\Sigma(Q)/d \Omega$; or 2)   $d\Sigma(Q)/d \Omega$ to  $G_{exp}(\delta_{SE})$. Another option is to simultaneously refine a SANS model against original SANS and SEMSANS data sets.

The first two options require a numerical Hankel transformation of experimental data.
The result of this transform is sensitive to the minimum and maximum  $Q$-(or $\delta_{SE}$)-values, to corrections for multiple scattering and incoherent scattering background, and to  the value of sample transmission (which is difficult to measure in the presence of strong forward scattering). 
  
The third option can be realized as a simultaneous refinement of a model against SANS data sets and the Hankel transform of the same model against SEMSANS data sets. The implementation of the key component of this option, namely, 
fitting of the Hankel transform of a SANS model to $V_s(\delta_{SE})$ measured in SEMSANS  is being implemented 
at Delft University of Technology in collaboration with developers of the SASview program \cite{SASVIEWSOFTWARE}. 

Please note that in the presence of very strong  small-angle scattering,  for example, from  large particles  and strong scattering contrast,  the scattering  may no longer be adequately modeled by  single-particle scattering models that are based on the first Born approximation\cite{Berk1988MSC}.  However, judging by the calculated $T_{SANS}$ and $T_{SEMSANS}$-values, this should not be  the case for all examples discussed above. 

%
%
\section*{\label{sec:conclusions}Summary}
 Based on the technical design of the SANS instrument SKADI, which will be built at ESS, and on preliminary characteristics of a SEMSANS add-on 
 we have shown that a wide range of length scales over 4 orders of magnitude, from $\approx$ 1 nm to $\approx$ 10 $\mu$m, can be covered simultaneously in a single measurement by combining the two techniques. 
The calculations involved several combinations of instrument parameters and wavelength ranges. The performance was discussed by considering specific examples from 
  soft matter, geoscience,  and advanced materials samples that were previously studied  with SANS,  USANS and imaging. 
  The results show that for all samples  adequate quality of SANS and SEMSANS measurements can be obtained  
   by choosing suitable parameters, such as the sample thickness.
Thus, a SEMSANS add-on on a SANS instrument such as SKADI  can be used to  simultaneously observe a wide range of length scales and for  time-resolved studies of kinetic processes in complex multiscale samples.

\section*{Acknowledgments}
This work was financially supported by Delft University of Technology and The Netherlands Organization for Scientific Research (NWO) through the OYSTER project as a contribution to the pre-construction phase of the  European Spallation Source. The authors thank J. Plomp as well as the ESS and SKADI teams for valuable discussions.

\section*{References}

\bibliography{mybib}


\appendix


\renewcommand{\theequation}{A.\arabic{equation}}
\setcounter{equation}{0}  
\section*{\label{sec:appendixlambda} Appendix A: An effect of  wavelength  resolution}
%
%

To estimate an effect of wavelength resolution on measured visibility, let us model the wavelength spread, $p(\lambda)$, by 
a gaussian with a standard deviation $\sigma_{\lambda}$ and 
the mean wavelength $\tilde{\lambda}$. Then, for  a given  pixel coordinate, $x$:
\begin{equation} 
\begin{split}
    \int p(\lambda) & \cos(2\pi x/\zeta) \mathrm{d} \lambda  =  \int p(\lambda) \cos(Cx\lambda) \mathrm{d} \lambda   = \\ & \cos(Cx\tilde{\lambda} )\exp(-(C x \sigma_{\lambda})^2/2)
\end{split}
  \label{eq:cosclambda}
\end{equation}
where  $C=  2c (B_2-B_1) \cot \theta_0 $  (cf. eq. \ref{eq:zeta}).
Thus, the reduction of oscillation amplitude is given by 
\begin{equation}
    R_{\Delta \lambda}(x) =\exp(-(C x\sigma_{\lambda} )^2/2)
  \label{eq:cosclambda1}
\end{equation}
and increases towards detector edges, i.e. towards larger $|x|$.

The usable detector width, $-x^{max}<x<x^{max}$, depends on the minimum acceptable $R_{\Delta \lambda}$-value, $R^{min}_{\Delta \lambda}$.
To observe at least one oscillation period,  $\zeta$ should not exceed $ 2 x^{max}$.
Using  eq. \ref{eq:spinecholengthSEMSANS}  this constraint  can be transformed into a constraint on the minimum observable spin-echo length:
\begin{equation}
    \delta^{min}_{SE~\Delta\lambda} =  \tilde{\lambda}d_1 /(2 x^{max)}
    \label{eq:constraint1b}
\end{equation}
where $x^{max}$ can be calculated from  eq. \ref{eq:cosclambda1} and  $R^{min}_{\Delta \lambda}$.

Specifically, if FWHM of a gaussian is 
$\Delta_{\lambda}=\tilde{\lambda} \mathrm{d} \lambda/\lambda$,
then,  $\sigma_{\lambda} =\Delta_{\lambda}/(2\sqrt{2 \ln 2})$, leading to  
\begin{equation}
    x^{max}=4\sqrt{- \mathrm{ln} (2) \mathrm{ln} (R^{min}_{\Delta \lambda})}/(C \tilde{\lambda}  (d \lambda/\lambda)) 
    \label{eq:cosclambda2}
\end{equation}
Then, from eqs.  \ref{eq:cosclambda2} and \ref{eq:constraint1b}:
\begin{equation}
     \delta^{min}_{SE~\Delta\lambda}= \delta_{SE} \frac{  \pi (d \lambda/\lambda)  }{4 \sqrt{- \mathrm{ln}(2)\mathrm{ln} (R^{min}_{\Delta \lambda})}  } 
    \label{eq:deltaSEmin}
\end{equation}
Thus, the constraint $\delta_{SE} \ge \delta^{min}_{SE~\Delta\lambda}$ is satisfied when 
\begin{equation}
d \lambda/\lambda \leq \frac{4 \sqrt{- \mathrm{ln}(2)\mathrm{ln} (R^{min}_{\Delta \lambda})}  }{\pi}
    \label{eq:deltaSEmin3}
\end{equation}


For SKADI,  $d \lambda/\lambda $ is expected 
to be in the range of 4\%-8\% and will certainly not exceed 20 \%  \cite{JakschBouwmanFrielinghaus2014SKADI}. From eq. \ref{eq:deltaSEmin3}, for $d \lambda/\lambda =0.2$, expected $R^{min}_{\Delta \lambda}$ is $0.965\approx 1$.
Thus, for this setup, finite wavelength resolution has no significant impact on the minimum $\delta_{SE}$ that can be achieved. 

Let us calculate the $x$ coordinate when the decrease in the oscillation amplitude becomes smaller than $R_{\Delta \lambda}$. 
From eqs.  \ref{eq:zeta}, \ref{eq:cosclambda1}, as well as from  $C$, $\sigma_{\lambda}$, and  $\Delta_{\lambda}$ defined above:
\begin{equation}
    x(R_{\Delta \lambda},\mathrm{d}\lambda/\lambda)=\frac{2 \zeta \sqrt{-\mathrm{ln} R_{\Delta \lambda} \mathrm{ln}2 }}{\pi (\mathrm{d}\lambda/\lambda )}
    \label{eq:distancedll}
\end{equation}
which can be rewritten as
\begin{equation}
    R_{\Delta \lambda}(x) =\exp(-( \frac{\pi\frac{\mathrm{d}\lambda}{\lambda} }{2}\frac{x }{\zeta})^2/\mathrm{ln}2)
    \label{eq:distancedll2}
\end{equation}

\renewcommand{\theequation}{B.\arabic{equation}}
\setcounter{equation}{0}  

\section*{\label{sec:appendixspatialdet} Appendix B: An effect of the  spatial detector resolution  on $\delta_{SE}^{max}$}

To estimate an effect of the spatial detector resolution on measured visibility,
we calculate the average over the width of one detector pixel, $p$:
\begin{equation}
    \frac{1}{p} \int_{\tilde{x}-p/2}^{\tilde{x}+p/2} \cos(\frac{2\pi}{\zeta}x) \mathrm{d} x = \cos( \frac{2\pi}{\zeta}\tilde{x}) \frac{\sin(\pi p /\zeta)}{p\pi/\zeta}
  \label{eq:cosy1}
\end{equation}
where $\tilde{x}$ is the coordinate of the center of the pixel.  
The reduction of oscillation amplitude  is
\begin{equation}
    R_{pixel} =\frac{\sin(\pi p /\zeta)}{p\pi/\zeta}\approx 1 - (p\pi/\zeta)^2/6 
  \label{eq:cosy2}
\end{equation}
Once  the minimum acceptable $R_{pixel}$ is set,   the minimum
 $\zeta$ can be found by numerical solution of eq. \ref{eq:cosy2} (the approximation in eq. \ref{eq:cosy2} is not fulfilled when $\pi p/\zeta$ gets large due to a small $\zeta$).
The maximum $\delta_{SE}$ due to a finite spatial resolution of SEMSANS detector is
\begin{equation}
    \delta^{max}_{SE~pixel}=\frac{\lambda d_1}{\zeta^{min}}
    \label{eq:deltaSEmax}
\end{equation}
 $\delta^{max}_{SE~pixel}$  given in Tab. \ref{table1SKADI} were calculated for $R_{pixel}^{min}=0.75$.

\renewcommand{\theequation}{C.\arabic{equation}}
\setcounter{equation}{0}  
\section*{\label{sec:appendixacceptance} Appendix C: An effect of acceptance angles on the normalized visibility }
In the absence of the sample, the intensity of spin-up (+) and spin-down (-) neutrons  as a function of detector pixel  coordinates $x$ and $y$  is an integral over all possible neutron trajectories:
\begin{align}
    & I^{\pm}_{se~0}(x,y)  =  \int \mathrm{d }s_x  \int \mathrm{d }s_y  \int  \mathrm{d } \theta_{i}  \int  \mathrm{d } \psi{i}   I_0(s_x, s_y,\theta_{i},\psi_{i}) \nonumber \\ 
    &  \times \frac{1 \pm P_0(s_x, s_y,\theta_{i},\psi_{i}) \cos( 2\pi x/\zeta  )}{2}
    \label{eq:acc1pre}
\end{align}
where $s_x$ and $s_y$ are the $x$- and $y$-coordinates  of a neutron at the  sample aperture    
and $\theta_{i}$ and $\psi_{i}$ are the incident angles in the $XZ$  and $YZ$-planes, respectively.  
This equation is a generalization of eq. \ref{eq:SEMSANSintensityNEW}, but 
 we neglect  effects of finite wavelength and detector resolution 
(cf. eq. \ref{eq:SEMSANSvis}).

For any  given pair of $x$ and $s_x$, and  $y$ and $s_y$, $\theta_{i}$ and $\psi_{i}$ are fixed by  definition:
\begin{equation}
    \theta_{i}=(x-s_x)/d_{1} \quad  \psi_{i}=(y-s_y)/d_{1}
    \label{eq:acc2}
\end{equation}
hence,  eq. \ref{eq:acc1pre} can be written as
\begin{align}
    I^{\pm}_{se~0}(x,y)  =  I_{0}^{sum}(x,y)/2 \pm I_{0}^{mod}(x,y)/2
    \label{eq:acc1general}
\end{align}
where 
\begin{equation}
     I^{sum}_0(x,y)  =  \int \mathrm{d }s_x  \int \mathrm{d}s_y I_0(s_x, s_y,\theta_{i},\psi_{i}) 
    \label{eq:acc1a}
\end{equation}
\begin{align}
    & I^{mod}_0(x,y)  = \int \mathrm{d }s_x  \int \mathrm{d}s_y I_0(s_x, s_y,\theta_{i},\psi_{i})P_0(s_x, s_y,\theta_{i},\psi_{i}) \\ &\times
    \cos( \frac{2\pi x}{\zeta} ) 
    \nonumber
    \label{eq:acc1b}
\end{align}
    The visibility, $V_0(x,y)$, is defined by (cf. eq. \ref{eq:SEMSANSintensityNEW}):
\begin{equation}
    V_0(x,y)\cos( 2\pi x/\zeta  )= I_{0}^{mod}(x,y)/I_{0}^{sum}(x,y)
    \label{eq:vis0}
\end{equation}

When a sample is present, the  intensity is 
\begin{equation}
    I^{\pm}_{se}(x,y)  =  T I^{\pm}_{se~0}(x,y) + I^{sum}(x,y)/ 2 \pm  I^{mod}(x,y) /2
    \label{eq:acc3}
\end{equation}
where the first term describes the contribution from direct beam, $T$ is the sample transmission defined in eq. \ref{eq:transtwoparts}.
Sample scattering comes in via:
%
\begin{align}
    &   I^{sum}(x,y) = \int  \mathrm{d } s_x \int  \mathrm{d } s_y \int   \mathrm{d } \theta_{i} \int \mathrm{d } \psi_{i}  S(s_x,x_y,\theta_{i},\theta_{f},\psi_{i},\psi_{f}) 
    \label{eq:acc3a} 
\end{align}
and, using $x=s_x +d_{1}\theta_{i} $ from eq. \ref{eq:acc2}:
\begin{align}
    & I^{mod}(x,y) = \int  \mathrm{d } s_x \int  \mathrm{d } s_y \int  \mathrm{d } \theta_{i} \int \mathrm{d } \psi_{i}  \cos(2\pi \frac{\theta_{i}d_{1}+s_x}{\zeta}  ) \nonumber \\ &  \times P_0(s_x, s_y,\theta_{i},\psi_{i})S(s_x,s_y,\theta_{i},\theta_{f},\psi_{i},\psi_{f}) 
    \label{eq:acc3b}
\end{align}
where $\theta_f$ and $\psi_f$ are the angles after scattering,\\  $S(s_x,s_y,\theta_{i},\theta_{f},\psi_{i},\psi_{f})$ is 
an effective scattering function which includes multiple scattering and sample attenuation and self-absorption,  it will be denoted as 
$S_{eff}$ for brevity.
The pixel coordinates $x$ and $y$ are related to the coordinates at the slit  via:
\begin{align}
    x= s_x+ d_{1}\tan \theta_f \approx s_x+ d_{1}\theta_i+d_{1}\theta_s \\
    y= s_y+ d_{1}\tan \psi_f \approx s_y+ d_{1}\psi_i+d_{1}\psi_s
    \label{eq:acc4}
\end{align}
where $\theta_s =\theta_f - \theta_i$ and $\psi_s=\psi_f - \psi_i$ are the scattering angles in the $XZ$ and $YZ $planes, respectively. 

From eq.    \ref{eq:acc4}, using $k_0=2\pi/\lambda$, $Q_x\approx  k_0\theta_s $, $Q_y\approx  k_0\psi_s $,
 $ \mathrm{d } \theta_{i} = \mathrm{d }Q_x/k_0$, $ \mathrm{d } \psi_{i} = \mathrm{d }Q_y/k_0$, 
and defining spin-echo length $\delta_{SE}=\lambda d_1/\zeta$,
    eq.  \ref{eq:acc3a} can be written as
\begin{align}
    &   I^{sum}(x,y) = \int  \mathrm{d } s_x \int \mathrm{d } s_y \int_0^{Q^{max}_x}  \mathrm{d } Q_x \int_0^{Q^{max}_y}  \mathrm{d } Q_y  \frac{ S_{eff} }{k_0^2} 
    \label{eq:acc4a} 
\end{align}
where $Q^{max}_x$ is a function of $x$ and $s_x$,  and $Q^{max}_y$ depends on $s_y$  and $y$. Further
\begin{align}
    & I^{mod}(x,y) = \cos(  \frac{ 2\pi x}{\zeta}  ) \int \mathrm{d } s_x \int \mathrm{d } s_y \int_{0}^{Q^{max}_x}  \mathrm{d } Q_x \int_0^{Q^{max}_y}  \mathrm{d } Q_y \nonumber \\ &
     \cos(Q_x \delta_{SE})\frac{S_{eff} }{k_0^2} P_0(s_x, s_y,\theta_{i},\psi_{i})  + \sin( \frac{2\pi  x}{\zeta}  )  \int  \mathrm{d } s_x \int  \mathrm{d } s_y \nonumber  \\  & \times \int_{0}^{Q^{max}_x}  \mathrm{d } Q_x \int_0^{Q^{max}_y}  \mathrm{d } Q_y 
     \sin(Q_x \delta_{SE})\frac{S_{eff}}{k_0^2} P_0(s_x, s_y,\theta_{i},\psi_{i})
    \label{eq:acc5}
\end{align}
Assuming  that
  $S_{eff}$ is symmetric with respect to $Q_x$,  the second integral in  eq. \ref{eq:acc5} is zero, which leads to
\begin{align}
    & I^{mod}(x,y) =  \cos( 2\pi \frac{x}{\zeta}  ) \int  \mathrm{d } s_x \int \mathrm{d } s_y   \int_{0}^{Q^{max}_x}  \mathrm{d } Q_x \int_0^{Q^{max}_y}  \mathrm{d } Q_y    \nonumber \\
     & \cos( Q_x \delta_{SE} )\frac{S_{eff}}{k_0^2}  P_0(s_x, s_y,\theta_{i},\psi_{i}) 
    \label{eq:acc5b}
\end{align}
The visibility, $V(x,y)$, is defined by
\begin{equation}
    V(x,y)\cos( 2\pi x/\zeta  )= \frac{ TI_{0}^{mod}(x,y) + I^{mod}(x,y)}{T I_{0}^{sum}(x,y) +  I^{sum}(x,y) }
    \label{eq:vis}
\end{equation}
Normalized visibility is given by:
\begin{equation}
    V_s(\delta_{SE})=\frac{V(x,y)}{V_0(x,y)}= \frac{ T + I^{mod}(x,y)/I_0^{mod}(x,y) }{T  + I^{sum}(x,y)/I_0^{sum}(x,y) }
    \label{eq:visnorm1}
\end{equation}
Here, the division by  $I_0^{sum}(x,y)$ is a normalization to the incident flux, 
and the division by $I_0^{mod}(x,y)$ normalizes to to the incident flux, and the empty beam polarization.

Thus,  the following two functions have to be obtained:
\begin{align}
    & \langle I^{sum}\rangle = \frac{ I^{sum}(x,y)} {I_0^{sum}(x,y)}  = \frac{2\pi}{k_0^2}\int_{0}^{Q^{max}}   S_{eff} Q \mathrm{d } Q
    \label{eq:n1} \\
    &  \langle  I^{mod} \rangle = \frac{ I^{mod}(x,y) }{I_0^{mod}(x,y)}= \frac{2\pi}{k_0^2} \int_{0}^{Q^{max}}    J_0(Q \delta_{SE}) S_{eff}Q \mathrm{d } Q 
    \label{eq:n2}
\end{align}
where,  for simplicity, we reduced a double integral over $Q_x$ and $Q_y$ to the integral over  $Q$ (under assumption of isotropic scattering and that
$Q^{max}= Q^{max}_x\approx Q^{max}_y$).

To calculate functions from eqs. \ref{eq:n1}, \ref{eq:n2}, we use the approach of Schelten and Schmatz and define two functions:
    $S(Q)$ which is $S_{eff}$ for a very thin sample,
and $H(Q)$ which is $S_{eff}$ for an arbitrary thickness $t$.
Eqs. 10-11 from Ref. \cite{ScheltenSchmatz1980} read  
\begin{eqnarray}
    s(\delta_{SE}) = {2\pi}\int_0^{\infty} J_0(Q\delta_{SE})  S(Q) Q \mathrm{d} Q \\
    \label{eq:ss1}
    h(\delta_{SE}) = {2\pi}\int_0^{\infty} J_0(Q\delta_{SE})  H(Q) Q \mathrm{d} Q 
    \label{eq:ss2}
\end{eqnarray} 
From eq. 12 in Ref. \cite{ScheltenSchmatz1980}:
\begin{equation}
    h(\delta_{SE}) = T k_0^2 \big[ \exp(s(\delta_{SE})/k_0^2) -1\big]
    \label{eq:ss3}
\end{equation}
where $T$ is the sample transmission.  $H(Q)$ is given by
\begin{equation}
    H(Q) =  \frac{1}{2\pi}\int_0^{\infty} J_0(Q\delta_{SE}) h(\delta_{SE}) \delta_{SE} \mathrm{d} \delta_{SE}
    \label{eq:ss4}
\end{equation}

We start by noting that $S(Q) = t \frac{d\Sigma(Q)}{d \Omega} $ 
and calculate  $s(\delta_{SE})$ from     eq. \ref{eq:ss1} as
\begin{equation}
    s(\delta_{SE})   = t{2\pi}\int_0^{\infty} J_0(Q\delta_{SE})  \frac{d\Sigma(Q)}{d \Omega} Q \mathrm{d} Q =G_{tot}(\delta_{SE})\xi_{tot}4\pi ^2 t
    \label{eq:ss5}
\end{equation} 
Then we calculate $h(\delta_{SE})$ from eq.  \ref{eq:ss3} and, after that, $S_{eff}=H(Q)$ from eq. \ref{eq:ss4}.
Finally, we calculate $\langle I^{sum}\rangle$ and $\langle I^{mod}\rangle$  from eqs. \ref{eq:n1},\ref{eq:n2} 
and use them to express $V_s(\delta_{SE})$ from  eq.  \ref{eq:visnorm1} as:
\begin{equation}
    V_s(\delta_{SE})  =    \frac{ T   + \frac{2\pi}{k_0^2} \int_{0}^{Q^{max}}    J_0(Q \delta_{SE}) H(Q)Q \mathrm{d } Q  }{T  + \frac{2\pi}{k_0^2} \int_{0}^{Q^{max}}     H(Q)Q \mathrm{d } Q }
    \label{eq:newapp2}
\end{equation}


Let us check the limit of eq. \ref{eq:visnorm1} as $Q^{max}\to \infty$. 
From eqs.  \ref{eq:n1}, \ref{eq:n2}, and \ref{eq:ss2} we arrive at
\begin{equation}
    V_s(\delta_{SE})  =    \frac{ T   +  h(\delta_{SE})/k_0^2  }{T  +  h(0)/k_0^2 }  
    \label{eq:newapp2limit1}
\end{equation}
and, using eq. \ref{eq:ss3} we arrive at
\begin{equation}
    V_s(\delta_{SE}) =   \exp([s(\delta_{SE}) - s(0)]/k_0^2)  
    \label{eq:newapp2limit2}
\end{equation}
which is the same as eq.  \ref{eq:pz}, as can be seen from eqs. \ref{eq:ss5}, and eqs. \ref{eq:defineksiexp}.,\ref{eq:GZfromIQhankelexp}.  

Please note that eq. \ref{eq:newapp2} also applies to the normalized polarisation measured in SESANS experiments.

Note also that the normalized visibility defined by eq.  \ref{eq:visnorm1} implicitly 
depends on $x$- and $y$-coordinates. Indeed, 
 $ \langle I^{sum}\rangle  $ and  $  \langle  I^{mod} \rangle $
    from eqs. \ref{eq:n1},\ref{eq:n2} depend on the integration range in $Q$ space, and this range depends on the range of accepted angles, and thus on $x$ and $y$. However, for the central part of the SEMSANS detector ($x \approx 0$, $y \approx 0$) such dependence is negligible, as can be seen from Appendix D.

\renewcommand{\theequation}{D.\arabic{equation}}
\renewcommand{\thefigure}{D.\arabic{figure}}
\setcounter{equation}{0}  
\setcounter{figure}{0}  

\section*{\label{sec:appendixangles}  Appendix D: The range of accepted scattering angles in SEMSANS}

\begin{figure}[htp]
\centering
\includegraphics[width=0.45\textwidth]{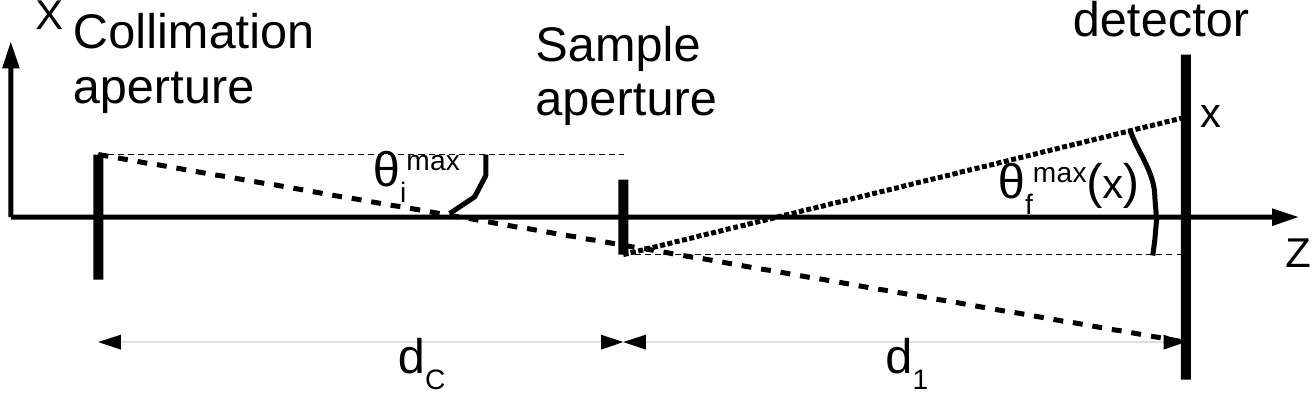}
    \caption{A sketch of geometrical constraints on incident angles ($\theta_i$) and final angles after scattering ($\theta_f$) for a pixel coordinate $x$.
    The  sizes of collimation and sample apertures along $X$ axis are $2S_C$ and $2S$, respectively.
}\label{fig:semsansacceptance}
\end{figure}

As can be seen from Fig. \ref{fig:semsansacceptance},  the  range of incident angles in the XZ plane is limited to $
\theta_{i}=\pm(S_C+S)/d_{C}$.
For the pixel coordinate $x$, the limits on  $\theta_{f}(x)$ are $(x\pm S)/d_{1}$.
Thus, the limits of $\theta_s(x)=\theta_f(x)-\theta_i$ are
\begin{equation}
    \theta_s(x) \in  ( \frac{x-S}{d_{1}} - \frac{S+S_C}{d_C} ; \frac{x+S}{d_{1}} + \frac{S+S_C}{d_C})
    \label{eq:acc7}
\end{equation}
The range of accepted scattering angles in the YZ plane, i.e. angles $\psi_s$, is defined  analogously. 

If measured "spin-up" and "spin-down" intensities, $ I^{\pm}_{se}(x,y)$, are summed up across the $Y$ axis of the SEMSANS detector, 
the range of accepted $\psi_s$ will be broader than the range of $\theta_s(x)$. As a result,  the assumption  $Q^{max}= Q^{max}_x\approx Q^{max}_y$ that was used 
to write eqs. \ref{eq:n1}-\ref{eq:n2} no longer holds. Consequently, in all equations starting with eqs.  \ref{eq:n1}-\ref{eq:n2} 
 the  integral over $|Q|$ has to be exchanged with a  double integral 
over  $Q_x$ and $Q_y$. 


\renewcommand{\theequation}{E.\arabic{equation}}
\setcounter{equation}{0}  

\section*{\label{sec:appendixtrans}  Appendix E: Sample transmission and  multiple scattering}

The fraction of the incident beam that did not interact with a sample  
is given by the sample transmission:
\begin{equation}
T(\lambda) = \exp(-[\Sigma_{a}(\lambda) +\Sigma_{coh}(\lambda)+\Sigma_{inc}(\lambda)]t)
    \label{eq:transtwoparts}
\end{equation}
where $t$ is sample thickness; $\Sigma_a(\lambda)$, $\Sigma_{coh}(\lambda)$,  and $\Sigma_{inc}(\lambda)$ are macroscopic absorption, coherent and incoherent scattering cross-section, respectively. 
$\Sigma_{inc}$ are assumed to be $\lambda$-independent and calculated from the atomic and isotopic sample composition using tabulated bound atom cross-sections. 
$\Sigma_{coh}(\lambda)$ is calculated from 
\begin{equation} 
    \Sigma_{coh}(\lambda)=  \int_{4\pi}  \frac{d\Sigma(Q)}{d \Omega}   \mathrm{d} \Omega = \frac{\lambda ^2}{2 \pi}\int_0^{Q^{max}}   \frac{d\Sigma(Q)}{d \Omega}   Q \mathrm{d} Q
    \label{eq:transcoh}
\end{equation}
where $Q=(4\pi/\lambda)\sin \theta/2$ and an isotropic scattering pattern is assumed. In general, $\frac{d\Sigma(Q)}{d \Omega}$ is the coherent differential scattering cross-section, and  $Q^{max}=4\pi/\lambda $. 
We neglected wide-angle coherent scattering and calculated $\frac{d\Sigma(Q)}{d \Omega}$ using a SANS model
and coherent bound scattering lengths.


For  $Q \geq$ 1 \AA\textsuperscript{-1}, a SANS model can not be expected to be reliable because the approximation of a continuous scattering length density distribution no longer holds. Therefore,  $Q^{max}$ is set to $4\pi/\lambda $ when  $4\pi/\lambda\leq $1 \AA\textsuperscript{-1}, and  1 \AA\textsuperscript{-1} otherwise.

To estimate the significance of  multiple small-angle scattering in the SANS experiment, we use   
\begin{equation}
    T_{SANS}(\lambda) = \exp(-\Sigma_{SANS}(\lambda)t) %
    \label{eq:transscat}
\end{equation}
where $\Sigma_{SANS}(\lambda)$ is calculated from eq. \ref{eq:transcoh}, 
with $Q^{max}$ set to   
$Q^{max}_{SANS}$  from eq. \ref{eq:qmaxsans} and the lower integration limit  is set to $Q^{min}_{SANS}$ from eq. \ref{eq:qminsans}.
Thus, only scattered neutrons that reach the two SANS detectors are taken into account.
Then,  the significance of  multiple  scattering is estimated as follows  \cite{SteinerLechner1991}.

If the probability to be scattered  once is $x$, twice  $x^2$ etc, then,
since $x<1$, the sum of all scattered neutrons is $x/(1-x)$. At the same time, it is just  $1-T_{SANS}$.
The result is $   x=\frac{1-T_{SANS}}{2-T_{SANS}}$ and 
the fraction of multiply scattered neutrons  to all scattered neutrons is $M=([x/(1-x)]-x)/[x/(1-x)]=x$.
For example, for $T_{SANS}=$  90\%,  $M$ is 9\%, for   $T_{SANS}=$ 80\%,  $M$ is 17\%.
Please note that this  is only a rough estimate  and that 
multiple small angle scattering will also be attenuated by absorption and incoherent scattering.

\end{document}